\documentclass{amsart}
\usepackage{epsf}
\usepackage{graphicx}
\newtheorem{lemma}{Lemma}[section]
\newtheorem{propos}[lemma]{Proposition}

\newtheorem{theorem}[lemma]{Theorem}

\newtheorem{corol}[lemma]{Corollary}
\theoremstyle{definition}

\theoremstyle{remark}

\newcommand{\CC}{\hbox{{$\mathcal C$}}}

\newcommand{\C}{\mathbb{C}}

\newcommand{\Z}{\mathbb{Z}}

\newcommand{\sign}{{\rm sign}}
\newcommand{\del}{\partial}

 \renewcommand{\span}{{\rm span}}

\newcommand{\extd}{{\rm d}}
\newcommand{\isom}{{\cong}}

\newcommand{\tens}{\mathop{\otimes}}
\newcommand{\la}{{\triangleright}}

\newcommand{\Ad}{{\rm Ad}}
\newcommand{\id}{{\rm id}}
\newcommand{\<}{\langle}
\renewcommand{\>}{\rangle}

\newcommand{\from}{{\longleftarrow}}
\newcommand{\note}[1]{}

\newcommand{\lbiprod}{{>\!\!\!\triangleleft\kern-.33em\cdot}}
\newcommand{\rbiprod}{{\cdot\kern-.33em\triangleright\!\!\!<}}

\newcommand{\eproof}{$\quad \diamond$\bigskip}

\newcommand{\eqn}[2]{\begin{equation}#2\label{#1}\end{equation}}

\newcommand{\done}{\overline{\partial}^1 \Theta_2}
\newcommand{\dtwo}{\overline{\partial}^2 \Theta_1}
\newcommand{\tone}{\bar\Theta_1}
\newcommand{\ttwo}{\bar\Theta_2}
\begin{document}



\title[Moduli of quantum Riemannian geometries]{\rm\large
MODULI OF QUANTUM RIEMANNIAN GEOMETRIES ON $\le 4$ POINTS}

\author{S. Majid + E. Raineri}%

\address{School of Mathematical Sciences\\
Queen Mary, University of London\\ 327 Mile End Rd,  London E1
4NS, UK}

\thanks{E.R. gratefully acknowledges the financial support of the United Kingdom Engineering and Physical Sciences Research Council and of the Italian Foundation "Angelo Della Riccia"}%



\maketitle

\begin{abstract} We classify parallelizable noncommutative manifold
structures on finite sets of small size in the general formalism
of framed quantum manifolds and vielbeins introduced in
\cite{Ma:rief}. The full moduli space is found for $\le 3$ points,
and a restricted moduli space for $4$ points. Generalised Levi-Civita connections
and their curvatures are found for a variety of models including models
of a discrete torus. The topological part 
of the moduli space is found for $\le 9$ points based on the known
atlas of regular graphs. 
\end{abstract}

\section{Introduction}

There has been a lot of interest
over the years \cite{DimMul, Bre, PasSit, Kra, BrzMa, MaSch, Ma:clif, MaRai:ele}
 in the specific application of noncommutative geometry\cite{Con}   to the commutative algebra
of functions on a finite set $\Sigma$ (usually a finite group) in
which the differential forms do not commute with functions. This
provides a systematic way of handling geometry on finite lattices
which, at the level of cohomology, electromagnetism and Yang-Mills
theory has already proven interesting and computable. Notably,
\cite{MaRai:ele} contains the moduli of $U(1)$-Yang-Mills on the
permutation group $S_3$ while \cite{Ma:clif} quantizes
$U(1)$-Yang-Mills theory on the finite group $\Z_2\times\Z_2$.

In this paper we want systematically to extend this theory to the
gravitational case. Some first steps are in \cite{Ma:rief}, to
which the present paper is a sequel. It was shown there that
finite groups have indeed a natural Riemannian geometry in a
vielbein and frame-bundle formalism\cite{Ma:rie} which was worked out in
detail for $S_3$ (it turns out to have ${\rm Ricci}$ essentially
proportional to the metric, i.e. an `Einstein manifold').
Similarly, the alternating group $A_4$ was considered in \cite{NML:rie}
and has an essentially unique invariant metric with 4-bein and an
associated spin connection with nonzero curvature but with {\rm
Ricci}=0, i.e. solves the vacuum Einstein equations. Hence the
system of equations for a framed quantum Riemannian manifold is
already known to have interesting nontrivial solutions. However,
for quantum gravity (or classical but finite gravity) we need a
better understanding of the moduli spaces of {\em all} metrics,
connections etc. and this is what we study now on small sets. Once one
has this, one can in principle begin to
quantize this moduli space in a path integral approach, i.e.
quantum gravity.

Section~2 starts with a brief account of the formalism for
algebras which we then rapidly specialise to the case
$\C(\Sigma)$, the algebra of functions in a finite set. That the
theory is a specialisation of a functorial construction that is
formulated for general algebras ensures that it is not ad-hoc
(indeed, this same theory can be specialised to classical geometry
and to q-deformed geometry for other choices of
algebra\cite{Ma:sph}). Following \cite{Ma:rief}, we find that for
finite sets $\Sigma$ the classification of `differential forms' or
exterior algebras of parallelizable type reduces to the
classification of finite regular graphs with vertices $\Sigma$ and
a fixed number $n$ arrows from every vertex. New results are
Theorem~2.1 showing in detail that the calculus is then inner, and
Theorem~2.2 for the construction of 2-forms. Both are needed in
the paper. Further ingredients in the formalism are a choice of
$n$-beins and a frame group $G$ (in our case a finite group)
acting on the vector space spanned by them. This gives the moduli
of `quantum framed manifold' structure on $\Sigma$. After this,
one may look for a compatible connection $\nabla$, find the
Riemann curvature and from this the Ricci tensors and Ricci
scalar. In this way we set up the theory that we are going to explore for
small numbers of points.

In Section 3.1 we analyze the case $\Sigma=\{x,y\}$ of two points
and frame groups $S_{2}, S_{3}$ acting on an einbein $ e_{1} $  parametrised by
a function $\Theta$. We find that
for each einbein there is a natural generalized Levi-Civita
connection
\[ \nabla(fe_{1} )=\extd f\tens e_{1}+2f\<\Theta\>e_{1}\tens e_{1} \]
for any function $f$, where $\<\ \>$ is the average value over the two points. This has
zero Riemannian curvature, which emerges as a typical feature on two points.
In our spin connection approach we find also the moduli of
 spin connections; for $S_{2 } $ framing we have a unique spin connection underlying $\nabla$. For $S_{3} $ we find a larger moduli of spin connections, with gauge curvature, underlying the Riemannian geometry itself (all giving the same $\nabla$ ).

In Section 3.2 we similarly cover the case  $\Sigma=\{x,y,z\}$ of
three points and frame groups $S_{2}, S_3$  acting on a zweibein. The
zweibein moduli space is itself nontrivial as an algebraic variety but we show
how put a generic point into a canonical form, and then study spin connections for
a fixed zweibein. A general feature for three points emerges, namely that in
all our models the Ricci scalar vanishes, but the Riemann and Ricci tensors themselves
generically do not. For $S_{2} $ we have a linear constraint on the zweibein to admit a connection, after which there is a 1-parameter family of connections. For $S_{3} $ there is no constraint on the zweibein and a 8-dimensional moduli of connections.

The canonical form for the vielbeins obtained in our analysis of 2 and 3 points in Section~3 is one where (after linear
transformations), one may restrict to vielbeins which have only a
scalar $\Theta_a$ associated to each edge. In Section~4 we
proceed to restrict attention to this canonical form, now for four point sets.
 Physically, the   modulus of
the veilbein assigns a `length' to each edge, while the natural
connectivity for 4 points is that of $\Z_2\times\Z_2$ (interpreted as a discrete model
of a torus), which we consider in Section~4.1 to~4.3; we consider various frame groups, among them an interesting
 choice (Section~4.2) is a frame group $\Z_4$ of `quarter rotations'  again as a discrete model of
 a torus; we find the most general connection, its Riemann and Ricci
curvatures, etc. This model has the feature (Theorem~4.4) that a fully metric compatible spin connection is determined uniquely by the zweibein, but with the latter further constrained. By contrast, our weaker 'skew metric compatible' or cotorsion free condition admits further parameters $a,b$ with the zweibein relatively unconstrained. We also see what happens if one takes too big a calculus on the frame group, namely additional unphysical modes emerge which do not, however, enter into the covariant derivative. This seems to us an important lesson for finite manifold-building by these methods. Section~4.4 completes the picture by covering the alternative connectivity of 4 points joined in a tetrahedron, which is
more like a sphere. Here with $\Z_3$ frame group of `one third' rotations we find an unusual but
interesting calculus, moduli etc without classical analogue.

Later, in Section~5  we make some first remarks on the quantum
theory, including a look at the discrete torus model on $\Z_{2}\times\Z_{2}$. Mainly, we find what we
show on this model to be a reasonable unitarity or *-structure on the system which
is needed to reduce the functional integrals to real variables.  We do not try to do the integrals themselves, which would be beyond the scope of our current analysis. 

Finally, Section~6 return to a more qualitative account of all
bidirectional framed geometries up to 9 points, deduced from the
known atlas of graphs\cite{ReaWil}. This covers the connectivity or
topological aspect of the vielbein moduli space. At this level a
vielbein amounts to a colouring of the graph into $n$-colours. For
each such vielbein, there are further continuous degrees of
freedom for matrices $e_a$ labelled according to the colouring $a$
(as seen in detail in Section~3). If we ignore these then we have a
in principle a `combinatorial quantum gravity' in which one sums over
all such colourings.

Let us note that `geometry' on finite sets in some form or other has a long pedigree.
Common to all approaches is the basic data of `differentials' as
defined by directed edges between vertices (a `digraph' or
quiver). Such objects are used in representation theory for
quivers formed on Dynkin diagrams. One also considers in that
context some kind of `vector bundles' with vector spaces over each
vertex albeit of varying dimension. Similarly in physics as well
as in simplicial cohomology one may `approximate' a manifold by a
finite triangulation and work on that. From the algebraic point of
view one does not actually need bidirectional edges, e.g. every
poset defines a connectivity graph and differential calculus with $x\to y$ if $x < y$ (albeit not a
parallelisable one if it is finite). This would be relevant to
modelling Lorentzian manifolds\cite{Sor} with $x\to y$ modeling a
time-like path from $x$ to $y$. Hence the deeper notions of
vielbeins and Riemannian geometry that we develop on such data
potentially has several applications.

\section{Preliminaries: formalism of quantum Riemannian manifolds}

Here we briefly recall the formalism of \cite{Ma:rief}. To tie in
with the general theory we start with a brief recap over general
algebras in Section~2.1. Then in Section~2.2 we specialise to the
finite set case in more detail than outlined in \cite{Ma:rief}. We
cover here only the parallelizable case where the frame bundle
algebra has a trivial tensor product form. There is a still more
general theory where the bundle is nontrivial, see \cite{Ma:rief},
but this needs much more machinery and we do not cover it here. It
would be needed for finite posets, for example.

\subsection{Over general algebras}

Let $M$ be a unital algebra. We equip $M$ with a differential
structure in the sense $(\Omega^1(M),\extd)$, where $\Omega^1(M)$
is an $M-M$ bimodule, and $\extd: M\to \Omega^1(M)$. This is a
notion common to all approaches to noncommutative geometry
including \cite{Con}. We also need $\Omega^2(M)$ or (in principle)
higher $\Omega^k(M)$ with $\extd^2=0$, for which we can take the
maximal prolongation of $\Omega^1(M)$ or any of its quotients.

In this context we define a (left) {\em vielbein} of $V$-bein as a
collection $\{e_a\}$ forming an $M$-basis of 1-forms
$e_a\in\Omega^1(M)$, i.e. $\Omega^1(M)\isom M\tens V$ where
$V=\span\{e_a\}$. One can also think equivalently of the $V$-bein
as a map $e:V\to \Omega^1(M)$ as in \cite{Ma:rief} if we regard
$V$ as a fixed abstract vector space. Given a vielbein we deduce
operators $\rho_a{}^b,\del^a:M\to M$ where \eqn{rhodel}{ e_a
f=\sum_b\rho_a{}^b(f)e_b,\quad \extd f=\sum_a(\del^a f)e_a,\quad
\forall f\in M} as an expression of the bimodule and exterior
derivative structure.

Next, we assume that we actually have an $A$-vielbein, i.e. we
require $V$ to be an $A$-comodule under a Hopf algebra $A$. There
is also a more general theory with $A$ merely a coalgebra, i.e.
this is not a critical assumption. We fix a left-covariant
differential structure $\Omega^1(A)$ on  the fiber of the frame
bundle. Like Lie groups, quantum groups are always parallelisable
and hence $\Omega^1(A)=A\tens \Lambda^1{}$ for some space of
invariant 1-forms $\Lambda^1{}$. This is a quotient of the
augmentation ideal $A_+$ of $A$ (classically it means the
functions that vanish at the group identity), i.e.
$\Lambda^1{}=A_+/Q_A$ for some left ideal $Q_A\subseteq A_+$. We
call $\Lambda^1{}^*\subset H_+$ the associated `quantum tangent
space', where we suppose $\Lambda^1{}$ is finite dimensional and
$H$ a Hopf algebra dually paired with $A$ (it plays the role
classically of the enveloping algebra of the Lie algebra of the
frame group). We will be interested only in the bicovariant case
as in \cite{Ma:rief} where one knows from the Worononwicz theory
\cite{Wor:dif} that $\Lambda^1{}$ is $\Ad$-stable or that
$\Lambda^1{}^*$ inherits $\Ad$ as a `quantum Lie bracket'. When
$A$ is coquasitriangular one knows that $\Lambda^1{}^*$ is in fact
a braided-Lie algebra \cite{GomMa:bra}. However, neither
assumption is critical for the geometry.

We let $\{f^i\}$ be a basis of $\Lambda^1{}^*$ and we denote by
$\la$ its left action inherited from the left action of $H$ on $V$ corresponding to the coaction
of $A$. It is only this action which is needed in the formulae
below. In this basis a spin connection means a collection of
1-forms $\{A_i\}$. Its torsion tensor corresponds to \eqn{tor}{
\extd e_a+\sum_i A_i\wedge f^i\la e_a} and we are interested in
torsion-free connections. We also (optionally) impose a regularity
or `differentiability' condition linking $\Omega^2(M)$ and
$\Omega^1(A)$, namely \eqn{regM}{ \sum_{ij}A_i\wedge A_j\,
\<f^if^j,q\>=0\quad \forall q\in Q_A.} This ensures that the
component 2-forms $\{F_i\}$ of the curvature
of the spin connection, namely
 \eqn{FM}{ F_i=\extd A_i+
\sum_{jk}c_i{}^{jk}A_j\wedge A_k} have a proper geometrical
interpretation as a curvature 2-form with values in
$\Lambda^1{}^*$. Here $c_i{}^{jk}=\<e_i,f^jf^k\>$ are structure
constants of the product of $H$ projected to $\Lambda^1{}^*$
(where $\{e_i\}$ is a dual basis of $\Lambda^1{}$).

For metrics we specialise to the case $g=\sum_{a,b}\eta^{ab}e_a
\tens_M e_b$ where $\eta\in V\tens V$ is a nondegenerate
$H$-invariant `local metric'. This is not the most general setup
up in \cite{Ma:rie}\cite{Ma:rief}, where one can consider $g$ an
arbitrary (but nondegenerate) 2-form. In our case the
cotorsion-free condition, which is the natural generalisation of
Levi-Civita metric compatibility in \cite{Ma:rie}\cite{Ma:rief},
is vanishing of \eqn{cotor}{ \extd e_a+\sum_i S^{-1}(f^i)\la
e_a\wedge A_i} where $S$ denotes the antipode of $H$.

Finally, we specialise to the case of $\Omega^2(M)$ constructed
from an $H$-equivariant projection $\pi:V\tens V\to V\tens V$
according to the scheme indicated in \cite{Ma:rief}. From the
above, we know that $\Omega^1(M)\tens_M\Omega^1(M)\isom M\tens
V\tens V$ allowing us to define surjections
\[\Omega^1(M)\tens_M\Omega^1(M)\to\Omega^2(M)\]
where we quotient out $M\tens \ker \pi$. In fact we define
$\Lambda$ as a quadratic algebra on $V$ with relations $\ker \pi$,
and $\Omega(M)\isom M\tens\Lambda$. Such a scheme imposes
constraints on $\pi$. In this setting there is a canonical lift
\eqn{lift}{ i:\Omega^2(M)\hookrightarrow
\Omega^1(M)\tens_M\Omega^1(M),\quad i(e_a\wedge e_b)=\pi(e_a\tens
e_b).}

Finally, we let $i(F_i)=\sum_{a,b}i(F_i)^{ab}e_a\tens_M e_b$
define the components in the $V$-bein basis of the lifted $F_i$.
Then \eqn{ricci}{ {\rm Ricci}=\sum_{i,a,b}i(F_i)^{ab}e_b\tens_M
f^i\la e_a.} The full Riemann curvature of the connection and the
covariant derivative acting on 1-forms are \eqn{rienabla}{ {\rm
Riemann}(\alpha)=\sum_{i,a}\alpha^a F_i\tens_M f^i\la e_a,\quad
\nabla \alpha =\sum_a\extd \alpha^a\tens_M e_a - \sum_{i,a}
\alpha^a A_i\tens_M f^i\la e_a} where $\alpha=\sum_a \alpha^a
e_a$. The derivation of these local formulae from a more abstract
theory is in \cite{Ma:rief}, in an equivalent comodule notation.

\subsection{Over finite sets}

We now specialise the above to the case $M=\C(\Sigma)$ where
$\Sigma$ is a finite set and $H=\C(G)$ where $G$ is a finite
group. In this case the possible $\Omega^1(\Sigma)$ are given by
subsets
\[ E\subset \Sigma\times\Sigma - {\rm diagonal}\]
of `allowed directions'. This is already known from \cite{Con} and
$E$ is the same as the structure of a quiver or digraph with
vertex set $\Sigma$ and the notation $x\to y$ whenever $(x,y)\in
E$. In the geometrical examples we typically expect $E$ symmetric
or `bidirectional' i.e. for every edge $x\to y$ there is an edge
$x\from y$, but we do not assume this in general. Explicitly,
\[ \Omega^1(\Sigma)=\span\{\delta_x\tens\delta y|\ x\to y\},
\quad \extd f=\sum_{y\in F_x}(f(y)-f(x))\delta_x\tens\delta_y;
\quad F_x=\{y|\ x\to y\}.\] \\
Note also that the $k$-fold product
\[ \Omega^1(\Sigma)\tens_M\cdots\tens_M\Omega^1(\Sigma)
=\span\{\delta_x\tens\delta_{x_1}\tens\cdots\tens\delta_{x_k}|\
x\to x_1\to x_2\to \cdots\to x_k\}\] i.e. the linear span of the
set of $k$-arcs. The bimodule structures are the pointwise ones
for products from the extreme left and right.

As explained in \cite{Ma:rief} a vielbein in this setting is
possible {\em iff} $E$ fibers over $\Sigma$ i.e. $F_x$ have
cardinality $n$ (say), independent of $x$. In this case an
$n$-bein is the specification of invertible $n\times n$ matrices
$e_{\cdot,x,\cdot}$ for each $x\in \Sigma$. Here $e_{axy}$ has
indices $a\in{1,\cdots,n}$ and $y\in F_x$. We write the inverses
as $e_a^{-1}{}^{xy}$ with \eqn{einvsig}{ \sum_{y\in
F_x}e_a^{-1}{}^{xy}e_{bxy}=\delta_{a,b},\quad \sum_a
e_a^{-1}{}^{xy}e_{axy'}=\delta^y_{y'}.} In this case the operators
(\ref{rhodel}) are \eqn{rhodelsig}{ \rho_a{}^b(f)(x)=\sum_{y\in
F_x}e_b^{-1}{}^{xy}f(y)e_{axy},\quad (\del^af)(x)=\sum_{y\in
F_x}(f(y)-f(x))e_a^{-1}{}^{xy}.}

A calculus on an algebra $M$ is inner if there is a 1-form $\theta$ with $\extd
f=[\theta,f]$ for $f\in M$.

\begin{theorem}cf\cite{Ma:rief} A finite set calculus equipped with
a vielbein is inner,
\[ \theta=\sum_a \Theta_a e_a,\quad \Theta_a(x)=\sum_{y\in F_x}e_a^{-1xy}.\]
Moreover, the maximal prolongation exterior algebra
$\Omega(\Sigma)$ has likewise $\extd=[\theta,\ \}$ (graded
anticommutator) and is generated by $\C(\Sigma)$ and the quadratic
algebra on the $\{e_a\}$ with relations
\[ \sum_{y\in F_{x,z}}\sum_{a,b}e_a^{-1xy}e_b^{-1yz}e_a\wedge e_b=0,
\quad \forall (x,z)\notin E\cup{\rm diag};\quad F_{x,z}=\{y|\ x\to
y\to z\}.\]
\end{theorem}
\proof  We define $\theta$ as stated. Then the explicit formulae
(\ref{rhodelsig}) allow one to verify that $\extd f=[\theta,f]$
for any function $f$, as required. The maximal prolongation of the
$\Omega^1$ is defined as the tensor algebra over $M=\C(\Sigma)$
modulo the relations in degree 2 imposed by extending $\extd$ as a
superderivation with $\extd^2=0$. More precisely, we lift any
1-form to the universal differential calculus over $\C(\Sigma)$,
apply the universal exterior derivative there, and then project
down to $\Omega^2$. That this should be well-defined defines the
minimal relations in degree 2 (which are the only ones imposed in
the maximal prolongation). In our case  as basis of the kernel of
the projection to $\Omega^1$ is given by $\delta_x\extd
\delta_z=0$ whenever $(x,z)\notin E\cup{\rm diag}$, so we require
for each such $(x,z)$ the relation
\[ \extd \delta_x \wedge \extd \delta_z=0.\]
We compute
\[ \extd \delta_x=\sum_a\sum_{y\in
F_\cdot}(\delta_x(y)-\delta_x)e_a^{-1\,\cdot\, y} e_a=\sum_a
(e_a^{-1\,\cdot,\ x}-\delta_x\Theta_a(x))e_a,\] where $\cdot$
denotes a functional dependence on points in $\Sigma$ and we adopt
the convention that $e_a^{-1 wx}=0=e_{awx}$ if $x\notin F_w$. Note
also that
\[ \sum_a e_a^{-1\,\cdot\, x}\rho_a{}^c(f)=\sum_{y\in F_\cdot} \sum_a e_a^{-1\,\cdot\,
x}e_c^{-1\,\cdot\, y}f(y)e_{\cdot\, y}=f(x) e_c^{-1\, \cdot\, x}\]
by(\ref{rhodelsig}) and (\ref{einvsig}). The latter also implies
that $\sum_a \Theta_a(x)e_{axy}=1$ if $y\in F_x$. Hence
\begin{eqnarray*}\extd \delta_x \wedge \extd
\delta_z&=&\sum_{a,b,c}e_a^{-1\, \cdot\, x}\rho_a{}^c(e_b^{-1\,
\cdot\ z})e_c\wedge e_b-\sum_{a,b,c}e_a^{-1\, \cdot\,
x}\Theta_b(z)\rho_a{}^c(\delta_z)e_c\wedge e_b\\
&&-\sum_{a,b,c}\delta_x\Theta_a(x)\rho_a{}^c(e_b^{-1\, \cdot\
z})e_c\wedge
e_b+\sum_{a,b,c}\delta_x\Theta_a(x)\Theta_b(z)\rho_a{}^c(\delta_z)e_c\wedge
e_b\\
&=&\sum_{b,c}e_c^{-1\, \cdot x}e_b^{-1 xz}e_c\wedge
e_b-\sum_{b,c}\delta_z(x)\Theta_b(z)e_c^{-1\,\cdot\, z}e_c\wedge e_b\\
&&-\delta_x \sum_{b,c}\sum_{y\in F_x}e_c^{-1xy}e_b^{-1yz}e_c\wedge
e_b+\delta_x\sum_{a,b,c}\Theta_a(x)\Theta_b(z)e_{axz}e_c^{-1xz}e_c\wedge
e_b.
\end{eqnarray*}
The first and last term vanish for $(x,z)\notin E$ and the second
term for $x\ne z$. Hence in this case we obtain precisely the
relation stated from the remaining third term. This completes the
proof of the result mentioned in \cite{Ma:rief}.

It is then a computation to write
\[ e_a=\sum_{(x,y)\in E} e_{axy} \delta_x \extd \delta_y
=\sum_{y\in F_\cdot} e_{a\, \cdot\, y}\extd \delta_y\] and obtain
$\extd e_a=\theta e_a + e_a \theta$. Note that the compatibility
of $\extd$ with the relations (\ref{rhodel}) for all $f$ more or
less requires this relation since applying $\extd$ to (\ref{rhodel})
gives $(\extd e_a-\{\theta,e_a\})f=\rho_a{}^b(f) (\extd
e_b-\{\theta,e_b\})$ after using (\ref{rhodel}) and that the calculus is
inner. \eproof

We note in passing that that by similar computations the maximal
prolongation has
\begin{eqnarray*}
\theta\wedge\theta&=&\sum_{a,b}\Theta_a e_a \Theta_b e_b
=\sum_{a,b,c}\Theta_a\sum_{y\in F_\cdot}e_c^{-1\, \cdot\, y}
\Theta_b(y) e_{a\, \cdot\, y}e_c\wedge e_b\\
&&=\sum_{b,c}\sum_{y\in F_\cdot}e_c^{-1\, \cdot\, y}
\Theta_b(y)e_c\wedge e_b= \sum_{a,b}\sum_{\cdot\, \to y\to
z}e_a^{-1\, \cdot\, y} e_b^{-1 yz}e_a\wedge e_b\end{eqnarray*}
which (in view of the relations for $\Omega^2(\Sigma)$) has
contributions only from $z=\cdot$ and $\cdot\, \to z$.  This is
not necessarily zero, i.e. $\theta$ is not necessarily closed
(rather, $\extd\theta=2\theta\wedge\theta$ so that
$\alpha=-2\theta$ is always a zero curvature $U(1)$ connection).

We also require for a $G$-covariant vielbein that $V=\span\{e_a\}$
is a $G$ module. The above constructions are all $G$-covariant
under these local transformations of $V$. To define more general
exterior algebras $\Omega(\Sigma)$ we let $\pi:V\tens V\to V\tens
V$ be a $G$-equivariant projection operator, with components
defined by $\pi_x(e_a\tens e_b)=\sum_{c,d}\pi_{ab}{}^{cd}e_c\tens
e_d$. We define operators \eqn{pixz}{ \pi_{x,z}:\C F_{x,z}\to \C
F_{x,z},\quad \pi_{x,z}{}^y{}_{y'}
=\sum_{a,b,c,d}\pi_{ab}{}^{cd}e_a^{-1xy}e_b^{-1yz}e_{cxy'}e_{dy'z}}
on the space spanned by 2-arcs with fixed endpoints $x,z$.

\begin{theorem} $\pi$ defines an exterior algebra with $\extd^2=0$
as a quotient of the tensor algebra on $V$ by the quadratic
relations
\[ \ker \pi=0\]
{\em iff}

${\rm (i)}\quad
\sum_{a,b,c,d}\pi_{ab}{}^{cd}e_a^{-1xy}e_b^{-1yz}e_{cxy'}e_{dy'z'}=0,
\quad \forall z\ne z';\ y\in F_{x,z},\ y'\in F_{x,z'}$

${\rm (ii)}\quad \sum_{y\in F_{x,z}}\pi_{x,z}{}^y{}_{y'}=0, \quad
\forall\ (x,z)\notin E\cup{\rm diag},\quad y'\in F_{x,z}.$
\end{theorem}
\proof We identify $\Omega^1(\Sigma)\tens_M\Omega^1(\Sigma)$ with
$\C(\Sigma)\tens V\tens V$ via the vielbein so that $\pi$ induces
left-module projection operators on this. These are therefore
given by  projection matrices $\pi_x$ on the space spanned by the
2-arcs from $x$, for each $x$. Their components are
\[ \pi_x{}^{yz}_{y'z'}=\sum_{a,b,c,d}\pi_{ab}{}^{cd}e_a^{-1xy}
e_b^{-1yz}e_{cxy'}e_{dy'z'}.\] We require that these are also
right module maps, which is the condition (i) stated. It means
that $\pi_x{}^{yz}_{y'z'}=\pi_{x,z}{}^y{}_{y'} \delta^{z'}{}_z$
for a family of projections $\pi_{x,z}$ for each fixed $x,z$.
These are the operators (\ref{pixz}). As explained in
\cite{Ma:rief} there is then a condition on the family of
projectors to ensure that the quotient
$\Omega^1(\Sigma)\tens_M\Omega^1(\Sigma)\to \Omega^2(\Sigma)$
factors through the maximal prolongation, namely the condition
(ii). This is necessary and sufficient for the relations in
$\Omega^2(\Sigma)$ defined by $\ker\pi$ to be compatible with the
extension of $\extd$ to 2-forms via the graded Leibniz rule.

\eproof

The maximal prolongation in Theorem~2.1 can be viewed as given by
a generalisation of this construction in which the projection
$\pi$ is allowed to vary from point to point, i.e. a field of
projections $\pi_x$. The more specific construction in Theorem~2.2
is necessarily a quotient of it by further relations.

Finally, we fix an $\Ad$-stable subset $\CC\subset G$ with
$e\notin \CC$ ($e$ here the group identity), e.g. a nontrivial
conjugacy class. These describe the bicovariant calculi
$\Omega^1(G)$ in the Woronowicz theory\cite{Wor:dif}. The space of
invariant forms $\Lambda^1{}$ in $\Omega^1(G)$ has basis $\{e_i|\
i\in\CC\}$. The dual basis of $\Lambda^1{}^*$ is $\{f^i\}$ with
$f^i=i-e$. The torsion and cotorsion equations then have the same
form
 (\ref{tor}) and (\ref{cotor}), with $Si=i^{-1}$ the
group algebra antipode. The regularity condition now reads
\eqn{regsig}{\sum_{ij=q}A_i\wedge A_j=0,\quad \forall q\notin
\CC\cup\{e\}.} This is empty if we chose the universal calculus on
$G$ (where
 $\CC=G-\{e\}$), but in general it is a quadratic constraint.
The curvature form is then \eqn{Fsig}{ F_i=\extd
A_i+\sum_{jk=i}A_j\wedge A_k-\{A_i,\sum_j A_j\}.} The formulae for
the {\rm Ricci} and {\rm Riemann} tensors and $\nabla$ have the
same form (\ref{rienabla}).

\section{Moduli of geometries on two or three points}

In this section we describe the moduli space of possible vielbeins
and metrics on 2 or 3 points, and moduli of spin connections and
their curvature for some points in the moduli of vielbeins with
respect to frame group $S_2$ or $S_3$.

More precisely, the moduli of possible vielbeins is in the first
place labelled by two natural numbers $m=|\Sigma|$ and $n$ a fixed
number of arcs from each point. For each $m,n$, the combinatorial
part of the moduli space consists of determining all possible
quiver structures with no self-arcs, i.e. all $E\subseteq
\Sigma\times\Sigma-{\rm diag}$ with $F_x$ of cardinality $n$ at
each $x\in \Sigma$. We interpret it as finding all possible
parallelizable $\Omega^1(\Sigma)$ with $n$-dimensional cotangent
space. Note that $\bar E$ where we flip the entries of $E$ defines
another calculus $\bar\Omega^1(\Sigma)$ and in the asymmetric case
one could (although we do not do it here) demand this to also be
parallelizable, with an associated number $\bar n$. There is a
corresponding moduli of geometries built on this arrow-reversed
calculus.

For $m=2$ or $\Sigma=\{x,y\}$ there is only one possibility,
namely $n=1$ and the quiver
\[  x\leftrightarrow y\]
up to relabellings. This is the universal calculus on $\Sigma$
where $E$ is as large as possible.

For $m=3$  or $\Sigma=\{x,y,z\}$  there are two cases for $n=1$,
namely
\[ {x\to y\atop {\nwarrow\ \swarrow\atop \displaystyle z}},
\quad x\leftrightarrow y\leftarrow z
\] up to relabellings. These are asymmetric. For $n=2$ there is only one
possibility, the universal calculus on $\Sigma$ again, which is
always symmetric. It is given by
\[ {x\leftrightarrow y\atop {\nwarrow\kern-8pt\searrow\ \swarrow\kern-8pt\nearrow\atop \displaystyle
z}}.\]

Next, for our projection matrix $\pi$ to define $\Omega^2(\Sigma)$
we make the `naive' choice \eqn{piflip}{ \pi={1\over 2} ( \id-\tau)} where
$\tau$ is the usual `flip' operator on the tensor product, i.e. we assume the basic 1-forms
anticommute. This
seems to give reasonable results for $n=2$  and a small number of points (in
general it would be too restrictive). For $n=1$ we choose $\pi=1$
(the choice $\pi=0$ is also allowed but not very interesting).
More generally, we should determine all possible equivariant
$\pi:V\tens V\to V\tens V$ for choice of frame group $G$ and a
representation $V$ of dimension $n$. The representation theory of
$G$ then dictates the possible equivariant projection matrices
$\pi:V\tens V\to V\tens V$. This is the representation theoretic
part of the moduli space.  In our case, we take symmetric groups
$S_2,S_3$ appropriate to a our small number of points. For $n=1$,
$V$ has to be trivial (we denote this by $\C$) or the sign
representation given by $(-1)^{l(g)}$ where $l$ is the length
function. For $n=2$ we have $V=\C\oplus \sign$,
$V=\sign\oplus\sign$ or, in the case of $S_3$ also its
2-dimensional representation. In all three cases $V\tens
V=\C\oplus \sign\oplus V$ and the `naive' $\pi$ (\ref{piflip})
projects out all but the sign representation here (cf in classical
geometry the top form transforms by the determinant under a linear
transformation). The invariant local metric $\eta$ up to a
normalisation is also classified by representation theory and we
take it as the generator of the natural trivial representation in
the decomposition of $V\tens V$.

Fixing all the above quasi-combinatorial data, we have a moduli
space \eqn{Veil}{ {\rm Vielbeins}_{m,n,E,\pi}=\{e_{a,x,y}\}/{\rm
GL_n}} consisting of $m$ $n\times n$ invertible matrices subject
to the constraints in Theorem~2.2. We divide by an overall $GL_n$
acting on the left and corresponding to a change of basis of $V$.
We arrive at a certain algebraic variety which we shall describe
first.

Finally, for a fixed vielbein and the above data, we look at the
moduli of spin connections for $\eta$. This last part requires us
to fix a differential structure on $G$. For $S_2$ the only choice
is the universal calculus $\Omega^1(S_2)$. For $S_3$ there is the
universal calculus and the calculus corresponding to the 2-cycles
conjugacy class. The remaining conjugacy class does not give a reasonable
geometry of $S_3$ (it is not connected) and does not appear to
give interesting results, so we omit it. In all cases we assume
that the action of $G$ on $V$ is not trivial when restricted to
the braided-Lie algebra generators $f^i$. Otherwise, they would
act as zero, the Riemann curvature would be automatically zero and
$\nabla$ would be just given by $\extd$ for any spin connection.
So we omit this uninteresting case in our analysis.

The case $\pi=0$ is trivial and we deal with it here. In this case
$\Omega^1(\Sigma)$ is the top degree so that there is no
constraint on the $\{e_{axy}\}$ other than being invertible. I.e.
\[ {\rm Vielbeins}_{m,n,E,0}=(GL_n)^{m-1}.\]
Similarly the torsion, cotorsion and regularity conditions are
empty and any collection of 1-forms $\{A_i\}$ are trivially a spin
connection, with zero curvature.

\subsection{Two points}

For $\Sigma=\{x,y\}$ the only choice is the universal calculus as
explained above, which has $n=1$, i.e. we look for a 1-bein $e_1$.
We write
\[ e_1=\alpha\delta_x\tens\delta_y + \beta\delta_y\tens\delta_x;
\quad e_{1xy}=\alpha,\quad e_{1yx}=\beta,\quad \alpha,\beta\ne
0.\] The partial derivatives and commutation relations are
\[ e_1f=\bar f e_1,\quad \del^1 f=(\bar f-f)\Theta;\quad
\Theta(x)=\alpha^{-1},\quad \Theta(y)=\beta^{-1};\]
\[ \bar f(x)=f(y), \quad \bar f(y)=f(x).\] The generating 1-form and
exterior derivative are
\[  \theta=\Theta e_1,\quad \extd f=(\bar f-f)\theta\]
For $\Omega^2(\Sigma)$ we have only one nontrivial possibility,
namely $\pi=1$, which gives the maximal prolongation with no
relations in the exterior algebra (it is the universal exterior
algebra on $\Sigma$). The conditions in Theorem~2.2 are empty as
the assumptions are never satisfied. Here $\Lambda=\C[e_1]$ and
each $\Omega^k(\Sigma)$ is 1-dimensional. The exterior derivatives
are defined by the graded-Leibniz rule and
\[ \extd e_1=(\Theta+\bar\Theta)e_1^2.\]
\begin{propos}
For any  $\Theta$ the dimensions of $H^i$ are $1:0:1$. Here \[H^0=\C.1,H^2=\C.\delta_{x} e_1^2\]
\end{propos}
\proof
First of all we show explicitly that, in accordance with theorem 2.2,
$d^2=0$.
\[d(df)=d(\Theta(\bar f -f))=\Theta \bar \Theta (f-\bar f) +\Theta \bar \Theta (\bar f-f)=0\]
The functions $f$ such that $df=0$ are the constant ones so the nullspace of $d$ acting on $\mathbb{C}[\Sigma]$ is 1-dimensional (and therefore, $p_0=1$). Since the dimension of
$\mathbb{C}[\Sigma]\otimes \Omega^1(\Sigma)=2$ this means that the image of d
in $\mathbb{C}[\Sigma]\otimes \Omega^1(\Sigma)$ is 1-dimensional.
If $\omega=f e_1$ is a one-form, the $d\omega=0$ if and only if
$\bar f \Theta+ f \bar \Theta=0$, or $f=\Theta (\delta_x-\delta_y)$
which implies that the nullspace of $d$ contained in
$\mathbb{C}[\Sigma]\otimes \Omega^1(\Sigma)$ is one dimensional.
Then, $p_1=1-1=0$. In turn, the image of $d$ in $\mathbb{C}\otimes \Omega^2(\Sigma)$ is one-dimensional, and so $p_2=2-1=1$
(every two form is in the kernel of $d$).$H^2$ is spanned over $\C$ by $\delta_x e_1^2$(or $\delta_y e_1 ^2$).
\eproof\\
Since we are working modulo an overall change of basis including
normalisation, only $\alpha^{-1}\beta$ is significant, so
\[ {\rm Vielbeins}_{2,1,{\rm univ},1}=\C^*\]

Next we look at spin connections. For group $G$ we assume a
symmetric group acting in the only nontrivial possibility, the
sign representation on $e_1$. Thus
 $f^i\la e_1=0$ if the permutation $i$ is
even and $f^i\la e_1=-2e_1$ if $i$ is odd. For $S_2$ we have only
the universal calculus, hence only one $f^i$ where $i=(12)$. We
write $A=a e_1$ for a function $a$. Then the torsion-free
condition becomes
\[ \Theta+\bar\Theta-2a=0\]
which is also the cotorsion-free condition, while the regularity
condition is empty. Hence for each 1-bein there is a unique spin
connection
\[ a={\alpha+\beta\over 2\alpha\beta}={\Theta+\bar\Theta\over 2}\]
which is a constant function. Its curvature is
\[ F=\extd A-2A^2=0\]
which means that the Riemann tensor is also zero. The covariant
derivative is \eqn{nab2S2}{ \nabla (fe_1)=\extd f\tens e_1+f
(\Theta+\bar\Theta)e_1\tens e_1.}

For $S_3$ we with its 3-dimensional (2-cycles) calculus we have
$f^i\la e_1=-2e_1$ and writing the three components functions
$a_1,a_2,a_3$ of the spin connections in directions

$(12),(23),(13)$, the torsion and cotorsion conditions for any
fixed 1-bein become
\[ \Theta+\bar\Theta-2(a_1+a_2+a_3)=0\]
while the regularity (which does not depend on the representation)
is
\[ a_1\bar a_2+a_2\bar a_3+a_3\bar a_1=0.\]
There are different classes of solutions including a 2-dimensional
part of the moduli space of spin connections for a generic 1-bein.
The curvature is
\[ F_i=(a_i-\bar a_i)(\bar\Theta-\Theta)e_1\wedge e_1\]
and is typically nonzero if the factor
$(\bar\Theta-\Theta)(x)={\alpha-\beta\over\alpha\beta}$ is
nonzero. On the other hand, we find (\ref{nab2S2}) again, with
zero Riemann curvature.

For $S_3$ with its 5-dimensional (universal) calculus, a spin
connection consists of components $b_1,b_2$ in the 3-cycles
directions which are unconstrained, and $a_1,a_2,a_3$ in the
2-cycles directions, with the single linear equation
\[ \Theta+\bar\Theta-2(a_1+a_2+a_3)=0\]
for vanishing of torsion and cotorsion. The regularity condition
is empty. So here the moduli space of connections is linear for
each vielbein. There is typically curvature at the frame bundle
level but again the Riemann curvature vanishes since $\nabla$ is
still given by (\ref{nab2S2}).

We conclude for 2 points that increasing the frame
braided Lie algebra allows more spin connections but these do not
enter into the Riemannian geometry itself. Instead, we find a
unique generalised Levi-Civita type covariant
derivative (\ref{nab2S2} )  for each einbein, and it has zero Riemannian curvature.

\subsection{Three points} For $\Sigma=\{x,y,z\}$ there are
two fibrations for $n=1$ and one for $n=2$ as explained above.

For $n=1$ a vielbein means three invertible numbers $\{e_{\cdot
x\cdot}\}, \{e_{\cdot y\cdot}\}, \{e_{\cdot z\cdot}\}$. However,
both types of fibrations for $n=1$ imply $\pi=0$ as the only
solution. This is forced by the conditions in Theorem~2.2 as
follows. For the triangular fibration the 2-arcs are \[ x\to y\to
z,\ y\to z\to x,\ z\to x\to y\] but then condition (ii) requires
$\pi_{x,z}{}^y{}_y=0$, which implies $\pi=0$. For the case of the
other fibration the 2-arcs are
\[ z\to x\to y,\ x\to y\to x,\ y\to x\to y.\]
In this case condition (ii) requires $\pi_{z,y}{}^x{}_x=0$ and
hence $\pi=0$. Hence for $n=1$ only the trivial case $\pi=0$
already covered in general above is allowed.

For $n=2$ we have only one fibration, which is the universal
$\Omega^1(\Sigma)$.  Then a vielbein means in the first place
three invertible matrices
\[ e_{\cdot x\cdot}=X,\quad e_{\cdot y\cdot}=Y,\quad e_{\cdot
z\cdot}=Z.\] Because of the cyclic nature of the graph, we label
the columns of $X$ at $y,z$, of $Y$ as $z,x$ and of $Z$ as $x,y$.
There are two types of 2-arcs, namely
\[ x\to y\to z, \quad x\to z\to y,\quad y\to x\to z,\quad y\to
z\to x\,\quad z\to x\to y,\quad z\to y\to x\] or
\[ x\to y\to x,\quad x\to z\to x,\quad y\to x\to y,\quad y\to z\to
y,\quad z\to x\to y,\quad z\to y\to x\] Finally, we take the
`naive' form (\ref{piflip}) for $\pi$.  The condition (ii) in
Theorem~2.2 is empty because $\Omega^1(\Sigma)$ is universal.
Condition (i) gives equations of the form
\[0=\pi^{xyz}_{xy'z'}={1\over 2} (e_1^{-1xy}e_2^{-1yz}-e_2^{-1xy}e_1^{-1yz})
(e_{1xy'}e_{2y'z'}-e_{2xy'}e_{1y'z'})\]
for $z'\ne z$ and $x\to y'\to z'$. Similarly for other 2-arcs in
place of $x\to y\to z$. The allowed cases are vanishing of
\[ \pi^{xyz}_{xzy},\quad \pi^{xyz}_{xyx},\quad
\pi^{xyx}_{xzy},\quad \pi^{xzx}_{xyz},\quad \pi^{xzx}_{xzy},\quad
\pi_{xyz}^{xzy},\quad \pi_{xyz}^{xyx},\quad \pi_{xyx}^{xzy},\quad
\pi_{xzx}^{xyz},\quad \pi_{xzx}^{xzy}
 \] and the cyclic rotations of $(xyz)$. Finally, keeping in mind
 the factorisation in the formula for $\pi$, we define
 \[
 f(x,y,z)=X_{1y}Y_{2z}-X_{2y}Y_{1z}=X_{11}Y_{21}-X_{21}Y_{11}\]
 \[
 f(x,y,x)=X_{1y}Y_{2x}-X_{2y}Y_{1x}=X_{11}Y_{22}-X_{21}Y_{12}\]
 etc. Here the first two entries of $f$ determine the matrices
 used, while the second two entries of $f$ label the indices on
 the matrices. Similarly, we define $\bar f(x,y,z)$ etc. in the same way but
 with $X^{-t}$, $Y^{-t}$, $Z^{-t}$ the inverse-transposed
 matrices.  With these notations we see that
 \[ {\rm Vielbeins}_{3,2,{\rm univ},{\rm flip}}\]
 is the variety consisting of 3 invertible matrices $X,Y,Z$ subject to the
 relations
 \[0=\bar f(x,y,z) f(x,z,y),\quad 0=\bar f(x,y,z) f(x,y,x),\quad
0=\bar f(x,y,x) f(x,z,y) \]
\[0=\bar f(x,z,x) f(x,y,z),\quad
0=\bar f (x,z,x) f(x,z,y),\quad   0=f(x,y,z) \bar f(x,z,y)\]
\[0=f(x,y,z) \bar f(x,y,x),\quad 0= f(x,y,x) \bar f(x,z,y)\]

\[0=f(x,z,x) \bar f(x,y,z),\quad 0=f (x,z,x) \bar
f(x,z,y)\] and their cyclic rotations of $(xyz)$, and modulo an
overall $GL_2$.

In principle this could have several cases depending on which
factor vanishes in each case. One special case is
\[ f(x,y,z)=0,\quad f(x,z,y)=0,\quad \bar f(x,y,z)=0,\quad \bar
f(x,z,y)=0\] and its cyclic rotations.  These equations reduce to
\[ X_{21}Y_{11}=X_{11}Y_{21},\quad
Y_{12}X_{22}=X_{12}Y_{22},\quad X_{11}Z_{21}=Z_{11}X_{21},\quad
X_{22}Z_{12}=X_{12}Z_{22}.\] Up to an overall $GL_2$, this
component of the moduli space of vielbeins has the general
solution
\[ X=\begin{pmatrix}\alpha_1&0\\
0&\alpha_2\end{pmatrix},\quad Y=\begin{pmatrix}\beta_1&0\\
0&\beta_2\end{pmatrix},\quad Z=\begin{pmatrix}\gamma_1&0\\
0&\gamma_2\end{pmatrix}\] modulo a remaining $\C^*\times \C^*$
(e.g. up to $GL_n$ one can assume $\alpha_1=\alpha_2=1$.) We
have\[ \Theta_a(x)=\alpha_a^{-1},\quad
\Theta_a(y)=\beta_a^{-1},\quad\Theta_a(z)=\gamma_a^{-1}\] and
\[e_a f=R_a(f)e_a,\quad  \del^a f=(R_a(f)-f)\Theta_a\]
where we identify $\Sigma$ with $\Z_3$ and use its addition law
according to the conventions above to define $R_a(f)=f((\ )+a)$.
For the exterior algebra, by our choice of $\pi$, the exterior
algebra relations are $e_1\wedge e_2=-e_2\wedge e_1$ and
$e_1^2=e_2^2=0$.
We are finally ready to look at compatible spin connections, which we
do for groups $S_2$ and then $S_3$, with their natural nontrivial representations and
calculi. Note that since we have already chosen a diagonal form of
the vielbein moduli space, different actions are not all
equivalent.
\begin{propos}
For any values of $\Theta_1,\Theta_2$ the dimensions $p_i$  of $H^i$ are $1:2:1$. Here \[H^0=\mathbb{C}.1,H^1=\C.<-\Theta_1 \delta_y e_1+ \Theta_2 \delta _z e_2,-\Theta_1 \delta_{x} e_1+\Theta_2 \delta_{y} e_2>,H^2=\C.e_1\wedge e_2\]
\end{propos}
\proof
Since in this case $R_2=R_1^{-1}$ we have
\[d(df)=(\Theta_1 R_1 \Theta_2 -\Theta_2 R_2 \Theta_1)(R_1 R_2 f-f)=0, \forall f \in \mathbb{C}[\Sigma]\]
and $d$ is cohomological.
By observing that $dc=0$ for any constant $c$, we have $p_0=1$ and the dimension of the image of $d$ in $\mathbb{C}[\Sigma]\otimes \Omega^1(\Sigma)$ (which is itself of dimension 6) is 2.
If $\omega=fe_1+ge_2$ is a one form, then $d\omega=0$ if and only if
$(-\Theta_2 \bar \partial^2 f+f \bar \partial^1 \Theta_2 +\Theta_1 \bar \partial ^1 g-g \bar \partial^2 \Theta_1)=0$.
This equation admits a 4 dimensional space of solutions, therefore $p_1=4-2=2$.
Moreover the image of $d$ in the two-forms is of dimension 2, and
given that $d$ sends every two-form to zero, one obtains $p_2=3-2=1$.
$H^1$ is spanned by $<-\Theta_1 \delta_y e_1+ \Theta_2 \delta _z e_2,-\Theta_1 \delta_{x} e_1+\Theta_2 \delta_{y} e_2>$, and $H^2=\C.e_1 \wedge e_2$
\eproof \\
For $S_2$ with its universal calculus, we choose the natural
action on $i=(12)$ on $V=\span\{e_1,e_2\}$ that flips the basis
vectors (hence by orientation reversal of the frame). The
invariant metric here is
\[ \eta=e_1\tens e_1+ e_2\tens e_2\] and the action of the braided-Lie
algebra generator of $S_2$ is $f^i\la e_1=e_2-e_1$ and $f^i\la
e_2=e_1-e_2$. Let us denote by $\bar\del^i\equiv R_i-\id$ the usual finite
difference on the group $\Z_3$, and $\<\ \>$ denotes the average value over
the three points.
\begin{propos} For 3 points, 2-dimensional cotangent space and $S_{2}$ frame group, existence of a torsion free cotorsion
free connection requires the zweibein to obey
\[ \Theta_1+R_1\Theta_2=\<\Theta_1+\Theta_2\>.\]
 In this case there is a 1-parameter family of connections of the form
\[ A=(\Theta_1-\lambda)e_{1}+(-R_{2}\Theta_{1}+\lambda  )e_{2} \]
for an arbitrary constant $\lambda$. The covariant derivative is
\[ \nabla e_{1}=-\nabla e_{2}=( (\Theta_{1}-\lambda)e_{1}+(-R_{2}\Theta_{1}
+\lambda)e_{2})\tens (e_{1}-e_{2}  ) \]
Its Riemannian and Ricci curvatures are
\[ {\rm Riemann} (e_1)=-{\rm
Riemann} (e_2)=\rho e_1\wedge e_2\tens (e_2-e_1),\ {\rm
Ricci}={\rho\over 2}(e_1+e_2)\tens(e_1-e_2).\]
where
\[ \rho=(2\lambda-\<\Theta_{1}+\Theta_{2}\>  )\bar\del^{2}\Theta_{1}.  \]
The Ricci scalar vanishes identically.
\end{propos}
\proof Writing a spin connection $A=a e_1 + b e_2$, the
torsion and cotorsion equations reduce to
\[ \bar\del^1\Theta_2=\bar\del^2\Theta_1=-(a+b)=R_2(a)+R_1(b)\]
and there is no regularity condition since the calculus on $S_2$
is universal. The third of these equations has
solution $b=-R_2(a)$ since $(\id+R_1)$ is invertible on $\Z_3$.
The full system for a vielbein and spin connection then reduces to
invertibility of the $\Theta_i$ values, the stated constraint on the zweibein, and
\[   a=\Theta_1 -\lambda,\quad b=-R_2\Theta_1+\lambda\] for an arbitrary
constant $\lambda$.  A straightforward computation then gives the curvature as
\[ F=\extd A-A^{2}=\rho e_1\wedge e_2 \]
for $\rho$ as stated, which Riemann  curvature as the action of $F$. One may also compute this directly from the covariant derivative stated. Finally, for the antisymmetrization projector that we use, the lifting map $i$ is
\eqn{iflip}{i(e_{1}\wedge e_{2})={1\over 2}(e_{1}\tens e_{2}-e_{2}\tens e_{1}).}
Using this to lift the 2-form values of the Riemann tensor and contracting as in (\ref{ricci} ) we obtain the Ricci tensor as stated. Its further contraction by the inverse metric is then zero. \eproof

  We see among other things that $\Theta_{2}$ is determined up to a constant from $\Theta_{1} $,
  i.e. not every zweibein is allowed. On the other hand, for a generic allowed
zweibein we have zero full curvature for a unique spin connection in the family, given
by $\lambda={1\over 2}\<\Theta_1+\Theta_2\>$.  Otherwise the curvatures are nonzero.

For $S_3$ with its standard 2-dimensional irreducible
representation and 2-cycles calculus, we have now
$i=(12),(23),(13)$ (as $i$ ranges 1,2,3) with the above flip
action of $(12)$ extended to a permutation of $e_1,e_2,e_3\equiv
-e_1-e_2$. The invariant metric is
\[ \eta=e_1\tens e_1+e_2\tens e_2+{1\over 2}(e_1\tens e_2+e_2\tens
e_1)\] and the action of $S_3$ on the vielbein is
\[
\begin{array}{lll}
f^1\triangleright e_1=e_2-e_1, & f^2 \triangleright e_1=0, & f^3
\triangleright e_1=-2e_1-e_2 \\
f^1\triangleright e_2=e_1-e_2,& f^2 \triangleright e_2=-e_1-2e_2, &
f^3 \triangleright e_2=0.
\end{array}
\]
\begin{propos} For 3 points, 2-dimensional cotangent space and $S_{3}$ frame group, the zweibein
is unconstrained and the torsion free cotorsion free connections are of the form
\[ A_{i}=a_{i}e_{1}+b_{i}e_{2};\quad a_{1}=a,\quad b_{1}=b\]
\[
 a_{2}={1\over 2} ( \bar\Theta_{1}-a),\quad b_{2}=R_{2}\bar\Theta_{1}+b,\quad a_{3}=R_{1}\bar\Theta_{2}+a,
\quad b_{3}={1\over 2}(\bar\Theta_{2}-b )   \]
for arbitary functions $a,b$ and costants $\lambda,\mu$. Here $\bar\Theta_{1}\equiv\Theta_{1} -\lambda $
and $\bar\Theta_{2}\equiv \Theta_{2}-\mu$ are notations. For a regular
connection we would need in addition:
\[ a R_1 b_2-b R_2 a_2+a_3 R_1 b-b_3 R_2 a+a_2 R_1 b_3 - b_2 R_2
a_3=0\]
\[ a_2 R_1 b-b_2 R_2 a+a R_1 b_3-b R_2 a_3+a_3 R_1 b_2-b_3 R_2
a=0.\]
The covariant derivative for the connection is
\[ \nabla e_1=(3a+2R_1 \bar\Theta_2)e_1 \otimes e_1 +R_1 \bar\Theta_2e_1 \otimes e_2 +\bar\Theta_2 e_2 \otimes e_1 +\frac{1}{2}(\bar\Theta_2-3b)e_2 \otimes e_2\]
\[
 \nabla e_2=\frac{1}{2}(\bar\Theta_1-3a)e_1 \otimes e_1 +\bar\Theta_2 e_1 \otimes e_2 +R_2 \bar\Theta_1 e_2 \otimes e_1 +(3b+2 R_2 \bar\Theta_1)e_2 \otimes e_2
\]
\end{propos}
\proof The torsion equations for a
spin connection with components $a_{i},b_{i} $  are
\[
\bar\del^1\Theta_2+a_1+b_1+2b_3-a_3=0,\quad
-\bar\del^2\Theta_1-a_1-b_1+b_2-2a_2=0\] and the cotorsion
equations
\[
 \bar\del^1\Theta_2-(R_2 a_1+R_1 b_1-R_2
a_3+2R_1 b_3)=0\]
\[-\bar\del^2\Theta_1+R_1 b_1+R_2 a_1-R_1 b_2+2R_2 a_2=0.\]
By combining these equations and using similar methods as in the
previous $S_{2} $ examples, one finds that their general solution is of the
form:
\[
 \Theta_1=2a_2+a_1+\lambda,\quad R_1(\Theta_2)=a_3-a_1+\mu\]
\[ b_1+2b_3=R_2(a_3-a_1),\quad 2b_3+b_2=R_2(2a_2+a_3)\]
for some constants $\lambda,\mu$. This means that for a fixed
vielbein and constants $\mu,\lambda$ the equations for a connection are solved as stated.
 One then writes out the covariant derivative and the optional regularity condition. \eproof

We can see here (and also in our previous examples) why full metric compatibility $\nabla\eta=0$ is too strong in finite noncommutative geometry (which is why we need our weaker cotorsion-free condition):

\begin{propos}  The covariant derivatives above do not fully preserve the metric unless $a=b=0$ and
$\Theta_{1}=\lambda, \Theta_{2}=\mu$  are constant.
\end{propos}
\proof We compute \begin{eqnarray*}
\nabla \eta=&(\frac{9}{2}a+4 R_1\bar\Theta_2 +\frac{\bar\Theta_1}{2}) e_1 \otimes e_1 \otimes e_1 + (2 R_1 \bar\Theta_2+\bar\Theta_1)e_1 \otimes e_1 \otimes e_2 \\
&+(2 R_1 \bar\Theta_2+\bar\Theta_1) e_1 \otimes e_2 \otimes e_1 +(2 \bar\Theta_2+R_2 \bar\Theta_1) e_2 \otimes e_1 \otimes e_1\\
&+ (2 \bar\Theta_1+R_1 \bar\Theta_2)e_1 \otimes e_2 \otimes e_2 +(\bar\Theta_2+2 R_2 \bar\Theta_1) e_2 \otimes e_1 \otimes e_2 \\
&+ (\bar\Theta_2+2 R_2\bar\Theta_1)e_2 \otimes e_2 \otimes e_1 +(\frac{9}{2}b+4 R_2\bar\Theta_1+\frac{\bar\Theta_2}{2}) e_2 \otimes e_2 \otimes e_2
\end{eqnarray*}
For this to be zero forces  $a=b=\bar\Theta_{1}=\bar\Theta_{2}=0$ which translates as stated
since $\lambda,\mu$ are abitrary. \eproof

 One may proceed to compute the curvatures etc. for a general solution. Here we present the results for the special case where the zweibein is  constant with $\Theta_{1}=\lambda$, $\Theta_{2}=\mu $ say, but the $a,b$ are artibrary, i.e. the flat background but not flat spin connection case .

 \begin{propos} For constant zweibein but $a,b$ arbitrary, the Riemann and Ricci
 curvatures take the form
 \[ {\rm Riemann}(e_1)=(3 \del^2 a- \del^1b)e_1\wedge e_2\tens e_1-2 \del^1 b e_1\wedge e_2\tens
e_2\]
\[ {\rm Riemann}(e_2)=-\del^2 a e_1\wedge e_2\tens
e_1+3 \del^1 b e_1\wedge e_2\tens e_2\]
\[ {\rm Ricci}= -{3\over 4} \del^2a e_1\tens e_1
+{3\over 2}\del^1b e_1\tens e_2 -
{1\over 2} (3 \del^2a-\del^1b)e_2\tens e_1 + \del^1 b
e_2\tens e_2.\]
The Ricci scalar vanishes identically. The regularity condition is
\[a R_1(b)=0.\]
\end{propos}
\proof   We compute the gauge curvature of the spin connection as
\begin{eqnarray*} F_1&=&(\mu\bar \del^2 a+
\lambda \bar \del^1 b)e_1\wedge e_2\nonumber\\
F_2&=&-({1\over 2}\mu \bar\del^2 a-\lambda \bar \del^1 b)e_1\wedge e_2\nonumber\\
F_3&=&-(-\mu\bar \del^2 a+\lambda\bar \del^1 b)e_1\wedge e_2.\end{eqnarray*}
Its action on the zweibein then determines the Riemann curvature as stated, using  (\ref{rienabla} ) . We recall that
$\del^{i}=\Theta_{i} \bar\del^{i}$  is the geometrical partial derivative defined by $\extd$ and we revert to this. We use the same lifting map as in Proposition~3.1 and (\ref{ricci} ) to find
\[ {\rm Ricci}={1\over 2}\begin{pmatrix}F_{2}-F_{1}& F_{1}+2F_{2}\\
-F_{1}-2F_{3}& F_{1}-F_{3}\end{pmatrix} \]
in the $e_{i} $  basis. This gives the result stated. Finally, note that the inverse of the matrix in $\eta$ is
\[ \eta^{-1}={4\over 3} \begin{pmatrix}1 & -{1\over 2}\\
 -{1\over 2}& 1\end{pmatrix}   \]
 in the dual basis and it is this which we use to contract against the Ricci tensor to obtain the
 Ricci scalar. Independently of the details of $F_{i}$, we have this as ${2\over 3}(F_2-F_1+F_1-F_3-\frac{1}{2}(F_1+2F_2-F_1-2F_3))=0$, i.e. vanishes identically
 . \eproof

The general case may be worked out in the same way: the formulae for the $F_{i}$ are rather more complicated functions of the $a,b,\lambda,\mu,\Theta_{i} $, but the other steps follow the same pattern. In particular, the Ricci tensor has the same asymmetric form and the Ricci scalar vanishes in general.
We see that with 3 points, the conditions with frame group $S_{2} $  are a little strong and constrain the zweibein, while with $S_{3}$ there are an abundance of spin connections compatible with any zweibein, namely $a,b$ arbitrary (and two further parameters which one might fix for example by $\lambda=\<\Theta_{1}\> $ and $\mu=\<\Theta_{2}\> $) and that in all cases with three points, the Ricci scalar vanishes.  Note that we have not covered it here, but one has a similar picture for $S_{3} $ with its universal calculus; then there are five 1-forms $A_i$ for the spin
connection with linear equations for the torsion and cotorsion
that prescribe the derivatives of $\Theta_1,\Theta_2$ in terms of
the fields, and an empty equation for regularity.

\section{Geometries on 4 points}

For four points we will not be fully general as above but restrict
to the more interesting class of models featuring already in our analysis for 2,3 points.
First of all, we shall focus on the case of all arrows
bidirectional, i.e. a symmetric subset $E$ to define the calculus.
This means for four points that we have (a) the square connectivity which is a
2-dimensional calculus or (b) the universal or 3-dimensional
calculus. We look mainly at the former since it has a clear geometrical
interpretation as the connectivity of a torus, namely in Sections 4.1-4.3. Indeed, this is the
natural calculus for the group $\Z_2\times\Z_2$ viewed as a
discrete model of a torus. Section~4.4 covers the alternative of the universal calculus
on the basis which has the connectivity of a tetrahedron or discrete model of a sphere.

Next, rather than the full analysis, we shall restrict attention
to the diagonal vielbeins given by scalars attached to the edges
as we deduced up to equivalence for the 3 points case above. These
scalars are our remaining continuous degrees of freedom and allow
our square to `pulse' by stretching or contracting edges. Such a
restricted class is interesting for any fixed combinatorics. On
the other hand, we will have more choices for the frame group and
its calculus, still giving several models.

\subsection{Discrete torus as base space} 

Thus, in this section, and the next two, we write the vertices as
$\Sigma=\{(0,0),(1,0),(0,1),(1,1)\}$, using an additive group
notation.\\
 Over each point we have a fiber
\[F_{(0,0)}=\{(1,0),(0,1)\},\ F_{(1,0)}=\{(0,0),(1,1)\},\ F_{(0,1)}=\{(0,0),(1,1)\},\ F_{(1,1)}=\{(1,0),(0,1)\}\]
of order 2. We fix the connectivity by identifying these fibers by
vielbeins of the diagonal form
\[e_{1,x,y}=\Theta^{-1}_1(x)\delta_{y-x,(1,0)},\quad e_{2,x,y}=\Theta^{-1}_2(x)\delta_{y-x,(0,1)}\]
for any two points $x,y\in \Sigma$. This is the natural vielbein
on $\Z_2\times \Z_2$ with additional continuous nowhere-zero
functional parameters $\Theta_i$. We have the picture
\[\epsfbox{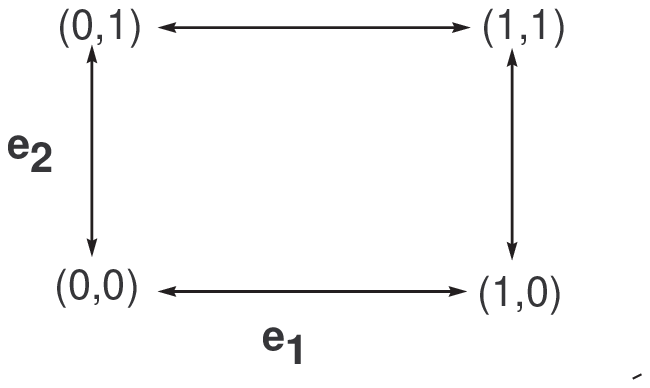}.\]
{}From each point in the lattice it is possible to move in two
directions, which correspond to the vectors $e_1$ and $e_2$. Here
$e_1$ translates adding the element $(1,0)$ of $\Z_2 \times \Z_2$,
and $e_2$ corresponds to moving by adding $(0,1)$. We define the
translation operators acting on the functions $f$ as
\[(R_1f)(x)=f(x+(1,0)),\quad (R_2f)(x)=f(x+(0,1))\]
and obtain the partial derivatives
\[
\partial^1f(x)=\left(f(x+(1,0))-f(x)\right)\Theta_1,\quad \partial^2f(x)=\left(f(x+(0,1))-f(x)\right)\Theta_2
\]
and commutation relations (which define the right multiplication
on $\Omega^1(M)$, left multiplication being the obvious one) as:
\[e_1f=R_1(f) e_1,\quad e_2f=R_2(f) e_2\]
for all functions $f$. This has the same form as we found up to
$GL_2$ for three points in the preceding section and completes our
description of $\Omega^1$ and its $\C[\Z_2\times \Z_2]$-basis
$\{e_1,e_2\}$.

 Next we fix the projector $\pi$ as the naive antisymmetrizer (\ref{piflip} ) on
the $e_i$ basis, again as we took in Section~3.2 for three points. We check
that for four points it obeys the condition needed in Theorem~2.2
to define the 2-forms $\Omega^2$. Thus, labelling the points as
$x,y,z,t$ where we start at $(0,0)$ and go around the square
clockwise, we have  the possible 2 arcs of two different forms:
\[\begin{array}{cc}
x\rightarrow y\rightarrow z  & y\rightarrow z\rightarrow t \\
x\rightarrow t\rightarrow z  & y\rightarrow x\rightarrow t \\
z\rightarrow t\rightarrow x  & t\rightarrow x\rightarrow y \\
z\rightarrow y\rightarrow x  & t\rightarrow z\rightarrow y \\
\end{array}
\]
or
\[\begin{array}{cc}
x\rightarrow y\rightarrow x & y\rightarrow z\rightarrow y\\
z\rightarrow y\rightarrow z & x\rightarrow t\rightarrow x\\
y\rightarrow x\rightarrow y & z\rightarrow t\rightarrow z \\
t\rightarrow x\rightarrow t & t\rightarrow z\rightarrow t
\end{array}
\]
to check out condition (i)in Theorem~2.2 we will consider the
two-arcs leaving from $x$, the starting point being irrelevant in
the reasoning which follows. This implies that we have to verify
the vanishing of ${\pi_x}_{yz}^{ yx},\ {\pi_x}_{yz}^{ tx},\
{\pi_x}_{yx}^{ yz},\ {\pi_x}_{tx}^{ yz}$ Now, in general
\[{\pi_x}_{y^{\prime} z^{\prime}}^{ yz}=(e_1^{-1xy}e_2^{-1yz}-e_2^{-1xy}e_1^{-1yz})(e_{1xy^{\prime}}e_{2y^{\prime}z}-e_{2xy^{\prime}}e_{1y^{\prime}z^{\prime}})\]
replacing in this expression the actual form of the arc, we
establish that \[{\pi_x}_{yz}^{ yx},\ {\pi_x}_{yz}^{ tx},\
{\pi_x}_{yx}^{yz},\ {\pi_x}_{tx}^{yz}\] are zero.
(We obtain a similar result swapping upper and lower indices).
The
second constraint of  Theorem~2.2, is not trivially satisfied in this case.
The conditions $\pi_{x,z}{}^y{}_{t}+\pi_{x,z}{}^t{}_{t}=0$ and $\pi_{x,z}{}^y{}_{y}+\pi_{x,z}{}^t{}_{y}=0$
both give
\eqn{constraint}{\Theta_1 R_1 \Theta_2=\Theta_2 R_2 \Theta_1,\quad {\rm i.e.},\quad \del_1\theta_2-\del_2\theta_1=0.}
Finally, we will take as a metric the element $\eta=e_1\otimes e_1
+ e_2\otimes e_2$  and we will take the lifting (\ref{iflip} )
 which is the natural choice for the antisymmetrizer
projector. The exterior differentials of the base elements are:
\[\extd e_1=\done e_1\wedge e_2 ,\quad \extd e_2=-\dtwo e_1\wedge e_2\]
where we recall that $\bar\del^{a}=\Theta_{a}^{-1}\del^{a}=R_{a}-\id $  are
the usual group finite differences.
\begin{propos}
For generic values of $\Theta_1, \Theta_2$  the dimensions $p_i$ of $H^i$ are  $1:2:1$. Here
\[H^0=\C.1,H^1=<(\Theta_2(y)\delta_t +\Theta_2(z)\delta_z )e_1,(\Theta_2(z)\delta_x+\Theta_2(y)\delta_t)e_2>,H^2=\C.e_1 \wedge e_2\]
\end{propos}
\proof
\[d(df)=(\Theta_1 R_1 \Theta_2 -\Theta_2 R_2 \Theta_1)(R_1 R_2 f-f)=0, \forall f \in \mathbb{C}[\Sigma]\]
which is zero due to the constraint we imposed on the $\Theta$s.
In the usual way, $p_0=1$, and the dimension of the image of $d$ in
$\mathbb{C}[\Sigma]\otimes \Omega^1(\Sigma)$ (itself of dimension 8) is 3.
A one form $fe_1+ge_2$ is in the nullspace of $d$ if and only if
it satisfies
$(-\Theta_2 \bar \partial^2 f+f \bar \partial^1 \Theta_2 +\Theta_1 \bar \partial ^1 g-g \bar \partial^2 \Theta_1)=0$; it's easy to find that the solution space of this equation has ,for generic values of $\Theta_i$ satisfying(~\ref{constraint}), dimension
5 (and therefore, $p_1=5-3=2$). This, in turn, implies the image of $d$ inside $\mathbb{C}[\Sigma]\otimes \Omega^2(\Sigma)$ is 3 dimensional.
Then, $p_2=4-3=1$. For generic values of $\Theta$, $e_1\wedge e_2$ is not in the image of
$d$, and gives a representative for $H^2$.
\eproof \\

For the remaining aspects of the geometry we fix the frame group
and its calculus. Then we can solve for the connections, curvature
etc. Even with all of the above choices, we have several models.

\subsection{Torus model with  $\Z_4 \subset SO(2)$ frame group}

Here we think of the additive group
$\Z_4=\{\overline{0},\overline{1},\overline{2},\overline{3}\}$ as
a discrete model of $SO(2)$, i.e. $90$-degree notations. This is
in keeping with our discrete model of a torus. We have on $\Z_4$
either the 3-dimensional universal calculus or the natural
2-dimensional calculus given by a square.

\subsubsection{3D calculus on $\Z_4$}

We choose the three-dimensional (universal) calculus defined on
$\Z_4$ by $\{\bar{1},\bar{2},\bar{3}\}$.
This corresponds to $f^{\bar{1}},f^{\bar{2}},f^{\bar{3}}$
acting as the corresponding rotation of the vielbein vectors $e_1$
and $e_2$, minus the identity, that is
\[\begin{array}{ll}
f^{\bar{1}}\triangleright e_1=e_2-e_1,& f^{\bar{1}} \triangleright e_2=-e_1-e_2\\
f^{\bar{2}}\triangleright e_1=-2e_1,& f^{\bar{2}}\triangleright e_2=-2e_2 \\
f^{\bar{3}}\triangleright e_1=-e_1-e_2, &
f^{\bar{3}}\triangleright e_2=e_1-e_2.
\end{array}
\]
\begin{propos}
The moduli space of torsion free cotorsion free connections on the
quantum Riemannian manifold above is given by:
\[
\begin{array}{l}
A_{\bar{1}}=\alpha e_1 +\beta e_2\\
A_{\bar{2}}=\frac{1}{2}(-\alpha+\beta-\gamma-\delta-\dtwo)e_1+\frac{1}{2}(-\alpha-\beta+\gamma-\delta-\done)e_2\\
A_{\bar{3}}=\gamma e_1 +\delta e_2
\end{array}
\]
for four functions $\alpha,\beta,\gamma,\delta$ subject to the
linear constraint
\[(R_1+R_2)a=0,\quad
(R_1+R_2)b=0\] where $a=\gamma-\alpha$ and $b=\beta-\delta$.
The corresponding covariant derivative is
\[\nabla e_1=(b-\dtwo)e_1 \tens e_1 +a e_1\tens e_2
+(a-\done) e_2 \tens e_1 -b e_2
\tens e_2\]
\[\nabla e_2=-a e_1 \tens e_1 +(b-\dtwo) e_1 \tens e_2
+b e_2 \tens e_1 +(a-\done) e_2
\tens e_2 \]
\end{propos}
\proof We want the connection to be torsion free, i.e., it has to satisfy
the following two linear equations:
\[
\begin{array}{l}
A_{1}^1 + A_{1}^2 + 2 A_{2}^2 - A_{3}^1 + A_{3}^2=-\done\\
-A_{1}^1 + A_{1}^2 - 2 A_{2}^1 -
A_{3}^1 - A_{3}^2=\dtwo
\end{array}
\]
from which we obtain the general solution above (without constraints on $\alpha,\beta,\gamma,\delta$.
In conformity to what we have done so far, we also demand that the cotorsion of the connection be zero. Notice that, differently from the previous cases investigated in this paper, the elements of the fibre group $\Z_4$ are not
of order 2, which implies that the action of $S$ on the  $f^i$s is not trivial; we have infact $S^{-1}(f^1)=f^3, S^{-1}(f^2)=f^2, S^{-1}(f^3)=f^1$,
and the zero-cotorsion condition can be put down (following \cite{cotor})
as
\[
\begin{array}{l}
-R_1 A_1^2+R_2A_1^1-2R_1A_2^2-R_2A_3^1-R_1A_3^2=-\done\\
R_1 A_1^2+R_2 A_1^1+2R_2 A_2^1-R_1 A_3^2+R_2A_3^1=\dtwo
\end{array}
\]
The requirement on the cotorsion translates into the constraint
$(R_1+R_2)a=(R_1+R_2)b=0$, where $a,b$ are defined in the statement of the proposition.
The regularity condition is empty, because the calculus on $\Z_4$ is the universal one.
We then compute the covariant derivative
using (\ref{rienabla}). \eproof \\
Note that the covariant derivative depends only on $a,b$ so these
parametrize the `physical' or effective moduli space, which is
therefore 4 dimensional: two functions on $\Z_2\times\Z_2$ modulo
the linear constraint. One may check that the torsion is indeed
zero, which is to say $\nabla \wedge e_1 =\extd e_1$ and $\nabla
\wedge e_2 = \extd e_2$. The cotorsion condition means
that $\nabla$ respects $\eta$ in a skew sense, as one may also
directly verify evaluating the cotorsion of the metric (when the
torsion is null) \[\Gamma \eta=(\nabla \wedge \id -\id \wedge
\nabla)(e_1 \otimes e_1+e_2 \otimes e_2)=0\] where
\[(\nabla \wedge \id)\eta=\done e_1 \wedge e_2 \otimes e_1-\dtwo e_1 \wedge e_2 \otimes e_2\]
and
\[(\id \wedge \nabla)\eta =(-(R_1+R_2)a+\done)e_1 \wedge e_2 \otimes e_1-((R_1 + R_2)b+\dtwo)e_1 \wedge e_2 \otimes e_2\]
(later on we will consider $\nabla\eta=0$ in the usual full
sense).
\begin{propos}
The Ricci scalar for the covariant derivative above is: \[R=\partial^1 b+\partial^2 a+\dtwo R_1 b+\done R_2 a-2b R_1 b-2 a R_2 a\]
\end{propos}
\proof: From the action of the $f^i$ we can then compute the
Riemann curvature using the general theory in Section~2, finding
now
\[\begin{array}{l}
{\rm Riemann}( e_1)=(-F_{1}-2F_{2}-F_{3})\tens e_1 +(F_{1}-F_{3}) \tens e_2 \\
{\rm Riemann}(e_2)=(-F_{1}+F_{3})\tens e_1 +
(-F_{1}-2F_{2}-F_{3})\tens e_2
\end{array}
\]
(Where we are denoting,for brevity, the coefficient functions of
$e_1\wedge e_2$ in the curvature components by $F_i$). In the same
way the Ricci tensor is:
\[{\rm Ricci}=\frac{1}{2}\left((F_{1}-F_{3})e_1\tens e_1 +(F_{1}+2F_{2} +F_{3})e_1 \tens e_2 -(F_{1}+2F_{2}+F_{3})e_2
\tens e_1+(F_{1}-F_{3}) e_2\tens e_2 \right),\] where
we identify the 2-forms $F_i$ with their scaler coefficients as multiples of the top form $e_1\wedge e_2$. and, taking the
trace in a standard way, the Ricci scalar is $R=F_{1}-F_{3}$ (this
and the other component $F_{1}+2F_{2}+F_{3}$ occur also in the
Riemann tensor so we see that the Ricci tensor vanishes if and
only if the entire Riemann tensor does). We can compute $F_{\bar
1},F_{\bar 2},F_{\bar 3}$ by means of:
\[
\begin{array}{l}
F_{\bar 1}=\extd A_1+A_2\wedge A_3+A_3\wedge A_2 -2 A_1\wedge A_1 -A_2\wedge A_1 -A_1\wedge A_2-A_3\wedge A_1- A_1\wedge A_3\\
F_{\bar 2}=\extd A_2+A_1 \wedge A_1 +A_3 \wedge  A_3 - A_1\wedge A_2- A_2\wedge A_1 -2A_2\wedge A_2 -A_3\wedge A_2 - A_2\wedge A_3\\
F_{\bar 3}=\extd A_3+A_2\wedge A_1+A_1 \wedge A_2 -A_3\wedge
A_1 -A_1 \wedge A_3-A_3 \wedge A_2 -A_2 \wedge A_3 -2 A_3 \wedge
A_3
\end{array}
\]
as inferred from (~\ref{Fsig}) where
\[
\begin{array}{ll}
\extd A_{\bar 1}=&(-\Theta_2 \bar \partial^2 \alpha+\alpha \done + \Theta_1 \bar \partial^1 \beta - \beta \dtwo)e_1\wedge e_2\\
\\
\extd A_{\bar 2}=&\frac{1}{2}\bigl(\Theta_1 \bar \partial^1(-\alpha-\beta+\gamma-\delta)+\Theta_2 \bar \partial^2(\alpha-\beta+\gamma+\delta)\\
&+\done(-\alpha+\beta-\gamma-\delta+2\Theta_1)+
\dtwo(\alpha+\beta-\gamma+\delta-2\Theta_2)\bigr) e_1 \wedge e_2\\
\\
\extd A_{\bar 3}=&(-\Theta_2 \bar \partial^2 \gamma+\gamma \done +
\Theta_1 \bar \partial^1 \delta - \delta \dtwo)e_1\wedge e_2
\end{array}
\]
and
\[\begin{array}{l}
A_{\bar 1}\wedge A_{\bar 1}=\alpha R_1 \beta-\beta R_2
\alpha\\
A_{\bar 1}\wedge A_{\bar 2}=\frac{\alpha}{2}(-R_1
\alpha-R_1 \beta+R_1 \gamma-R_1 \delta+\done)+\frac{\beta}{2}(R_2
\alpha-R_2 \beta+R_2 \gamma+R_2 \delta-\dtwo)\\
\end{array}
\]
etc. The detailed form of the curvature two-form is 
\begin{eqnarray*}
F_{\bar{1}}&=&-\partial^2 \alpha+\partial^1 \beta +\alpha(-R_1 \beta-R_1 \delta-R_2 \alpha+R_2 \gamma+\frac{\done}{2})\\
&&+\beta(R_2 \alpha+R_2 \gamma+R_2 \beta- R_2 \delta -\frac{\dtwo}{2})+\gamma(-R_1 \delta-R_1 \beta-R_1 \alpha+R_1 \gamma+\frac{\done}{2})\\
&&+\delta(R_1 \beta-R_1 \delta+R_2 \gamma+R_2 \alpha-\frac{\dtwo}{2})+\frac{\dtwo}{2}R_1(\beta-\delta)-\frac{\done}{2}R_2(\alpha-\gamma)\\
F_{\bar{2}}&=&\frac{1}{2}\partial^1(-\alpha-\beta+\gamma-\delta)+\frac{1}{2}\partial^2(\alpha-\beta+\gamma+\delta)\\
&&+\frac{\done}{2}(-R_2 (\beta-\delta)-\alpha+\beta-\gamma-\delta)+\frac{\dtwo}{2}(-R_1 (\alpha-\gamma)+\alpha+\beta-\gamma+\delta)\\
&&+\frac{\alpha}{2}(3 R_1 \beta-R_2 (\beta-\delta)-\dtwo+R_1 \delta)+\frac{\beta}{2}(-3 R_2 \alpha+ R_1 (\alpha-\gamma)-\done-R_2 \gamma)\\
&&+\frac{\gamma}{2}(3 R_1 \delta+R_2 (\beta-\delta)+\dtwo+R_1 \beta)+\frac{\delta}{2}(-3 R_2 \gamma-R_1 a+\done-R_2 \alpha)\\
F_{\bar{3}}&=&-\partial^2 \gamma+\partial^1 \delta+\frac{\done}{2}R_2 (\alpha-\gamma)-\frac{\dtwo}{2}R_1 (\beta-\delta)\\
&&+\alpha(-R_1 \delta-R_1 \alpha+R_1 \gamma-R_1 \beta+\frac{\done}{2})\\
&&+\beta(R_2 \alpha+R_2 \gamma +R_1 \beta- R_1 \delta-\frac{\dtwo}{2})+\gamma(-R_1 \delta+R_1 \alpha-R_1 \gamma+\frac{\done}{2}-R_1 \beta)\\
&&+\delta(R_2 \gamma +R_2 \alpha+R_2 \beta-R_2 \delta-\frac{\dtwo}{2})
\end{eqnarray*}
from which we compute the Ricci curvature etc. as above, and write in terms of $a,b$. 
 \eproof 

It is useful to observe that it's not mandatory to compute the curvature two-form in order to
get hold of the Riemann tensor.
One could also\cite{Ma:rief}  use the formula ${\rm Riemann}(e_1)=((\id \wedge \nabla)-(\extd \otimes \id))\circ \nabla (e_1)$ and similarly for $e_2$, which provides a useful check. 
Either way, the Riemann tensor turns out to have the form
\[{\rm Riemann}(e_1)=\rho e_1 \wedge e_2 \otimes e_1 +R e_1 \wedge e_2 \otimes e_2
,{\rm Riemann}(e_2)= -R e_1 \wedge e_2 \otimes e_1+\rho e_1 \wedge e_2 \otimes e_2\]
with \[\rho=\partial^2 b-\partial^1 a+2b R_1a-2a R_2b -\dtwo R_1 a+\done R_2 b\]
and $R$ the Ricci scaler computed above.  We see in particular that $a=b=0$ is a natural point in the
effective moduli space where the Ricci tensor (and the entire curvature) is zero. 

Next we consider full metric compatibility as opposed to the
weaker cotorsion condition.
\begin{theorem}
The metric $\eta$ satisfies the equation $\nabla \eta=0$, if and
only if
\[
a=\done,\quad  b=\dtwo ,\quad \bar \partial^2 (\bar \partial^1
\Theta_2)=0,\quad \bar \partial^1 (\bar \partial^2 \Theta_1)=0 
\]
The Ricci scalar is given by
\[R=-(\done)^2-(\dtwo)^2\]
\end{theorem}
\proof We only need to state explicitly the equality $\nabla
\eta=0$, as in:
\[\begin{array}{ll}
\nabla (e_1\tens e_1+e_2\tens e_2)=&2\bigl((b-\dtwo)e_1\tens e_1 \tens e_1+(a-\done)e_2 \tens e_1 \tens e_1\\
&+(b-\dtwo)e_1 \tens e_2 \tens e_2+(a-\done)e_2\tens e_2 \tens e_2\bigr) =0
\end{array}
\]
the solution to the above equation is $a=\done$,
$b=\dtwo$. The kernel constraint on $a,b$ then
requires the constraint on the vielbein. \eproof\\
 
 We see that not
every vielbein admits a strictly metric compatible condition -- in
general we need our weaker cotorsion-free condition. However, when
it does so, the covariant derivative is uniquely determined as in
classical Riemannian geometry.

\subsubsection{2D calculus on $\Z_4$}

We also consider the 2D calculus on $\Z_4$, defined by $\{\bar{1},\bar{3}\}$ with  $f^{\bar{1}}$ and $f^{\bar{3}}$,
acting as before. Our interesting result is that the geometric content is the same as the universal calculus above except that some redundant modes in the universal case are not present,  but replaced by a quadratic regularity condition.  

\begin{propos} With the above specification for the action, the moduli space of
torsion free, cotorsion free connections is given by:
\[
\begin{array}{l}
A_{\bar{1}}=(-\alpha-\frac{\dtwo}{2})e_1+(\beta-\frac{\done}{2}) e_2\\
A_{\bar{3}}=(\beta-\frac{\dtwo}{2}) e_1+ (\alpha-\frac{\done}{2}) e_2
\end{array}
\]
with the conditions \[(R_1+R_2)a=0,(R_1+R_2)b=0\]
where $a=\alpha+\beta$ and $b=\beta-\alpha$.
In terms of $a,b$ the covariant
derivative $\nabla$ is as before, in Proposition~4.1, and the regularity
condition reads
\[
 \dtwo\bar\del^2 a-\done\bar\del^1 b=0\]
\end{propos}
\proof Here the parameters $\alpha,\beta$ are not the same as in the previous section (but related to them). We solve the zero torsion condition
\[
\begin{array}{l}
A_1^1 + A_1^2 - A_3^1 + A_3^2=-\done\\
-A_1^1 + A_1^2 - A_3^1 - A_3^2=\dtwo
\end{array}
\]
which gives the solution above in terms of $\alpha,\beta$ or the combinations $a,b$,  but  free of any constraint on  the $a,b$. Next  we require the connection to have zero cotorsion:
\[
\begin{array}{l}
-R_1A_1^2 - R_1A_3^2 - R_2A_1^1 + R_2A_3^1=-\done\\
-R_1A_1^2 + R_1A_3^2 + R_2A_1^1 + R_2A_3^1=\dtwo
\end{array}
\]
and obtain the constraint $(R_1+R_2)a=(R_1+R_2)b=0$. 
We then compute the covariant derivative using the action of $f^{\bar{1}}, f^{\bar{3}}$. The regularity condition in this
case is given by:
\[A_{\bar{1}} \wedge A_{\bar{1}} +A_{\bar{3}} \wedge A_{\bar{3}}=0\]
\eproof
\\
\begin{corol}
The Riemann and Ricci tensors corresponding to the connection above have the  form (in terms of $a$ and $b$) as in Proposition~4.2
\end{corol}
\proof    This follows since the Riemann and Ricci tensors are determined by $\nabla$ which has the same form. It is also instructive (but a different computation) to compute them directly; as usual from the definition of the curvature
\[\begin{array}{l}
F_1=\extd A_1- 2A_1 \wedge A_1-A_3 \wedge A_1 -A_1 \wedge A_3\\
F_3=\extd A_3-A_1 \wedge A_3-A_3\wedge A_1 -2 A_3 \wedge A_3
\end{array}
\]
we compute the expression for the Riemann tensor:
\[
\begin{array}{l}
{\rm Riemann}(e_1)=(-F_1-F_3)\tens e_1 +(F_1 -F_3) \tens e_2\\
{\rm Riemann}(e_2)=(-F_1+F_3)\tens e_1 +(-F_1-F_3)\tens e_2
\end{array}
\]
inserting the actual form of $F_1$ and $F_3$ and the regularity condition.
\\
The Ricci tensor is  
\[\begin{array}{l}
{\rm Ricci}=\frac{1}{2}((F_1-F_3)e_1\tens e_1 +(F_1+F_3)e_1\tens e_2\\
\quad\quad\quad +(-F_1-F_3)e_2 \tens e_1 +(F_1-F_3)e_2 \tens e_2)\end{array}\]
\eproof \\

If we want Ricci flatness, we must force $F_1=F_3=0$.
Note that if the Ricci flat is null, so is the Riemann tensor.

The condition for the metric compatibility is the same as in Proposition~4.4
\begin{propos}
The metric $\eta$ satisfies the equation $\nabla \eta=0$, if and
only if
\[
a=\done,\quad  b=\dtwo ,\quad \bar \partial^2 (\bar \partial^1
\Theta_2)=0,,\quad \bar \partial^1 (\bar \partial^2 \Theta_1)=0.
\]
The regularity condition holds and the Riemann and Ricci tensors are as in Theorem~4.4
\end{propos}
\proof
We impose the condition $\nabla \eta=0$, which has the same shape as in the previous case. We then check that the regularity condition in Proposition~4.1 indeed holds for these $a,b$. 
\eproof

We conclude that moving to the 2D calculus on $\Z_{4}$  gives essentially the same
Riemannian geometry as using the 3D calculus but without some of the superfluous modes
that we found there. Instead, these are replaced by a regularity condition. This gives us some
insight into the 'correct' choice of calculus for the frame group and what happens if one
chooses one that is too big.

\subsection{Torus model with translations $\Z_2\times\Z_2$ as frame group}
We take now the frame group to be $\Z_2\times\Z_2$ acting by
`translation' on our base space which we recall is also the group
$\Z_2\times \Z_2$. We write the frame group elements as
$\bar{00},\bar{01},\bar{10},\bar{11}$, say. As before, we have two
choices for the calculus on the frame group.

\subsubsection{3D calculus on $\Z_2\times\Z_2$}
This is the universal calculus defined by $\{\bar{10},\bar{10},\bar{11}\}$
The corresponding $f$s act by
\[
\begin{array}{ll}
f^{\bar{10}}\la e_1=-2e_1,  & f^{\bar{10}}\la e_2= 0\\
f^{\bar{01}}\la e_1=0,      & f^{\bar{01}}\la e_2=-2e_2\\
f^{\bar{11}}\la e_1= -2e_1, & f^{\bar{11}}\la e_2= -2e_2
\end{array}
\]
\begin{propos}
The moduli space of torsion free, cotorsion free connections is
given by:
\[
\begin{array}{l}
A_{\bar{10}}=\alpha e_1- (\delta+\frac{\done}{2})e_2\\
A_{\bar{01}}=-(\gamma+\frac{\dtwo}{2}) e_1 +\beta e_2\\
A_{\bar{11}}=\gamma e_1 +\delta e_2
\end{array}
\]
We will use the ansatz $a=\alpha+\gamma,b=\beta+\delta$.
The
covariant derivative corresponding to this connection is:
\[
\nabla e_1=2a e_1 \tens e_1 -\done e_2 \tens e_1,\
\nabla e_2=-\dtwo e_1 \tens e_2 +2 b e_2 \tens e_2
\]
\end{propos}
\proof We solve the torsion condition
\[
\begin{array}{l}
2 A_{\bar{10}}^2 + 2 A_{\bar{11}}^2=-\done,\quad -2 A_{\bar{01}}^1 - 2 A_{\bar{11}}^1=\dtwo
\end{array}
\]
and the zero cotorsion condition
\[
\begin{array}{l}
-2 R_1A_{\bar{10}}^2 - 2 R_1A_{\bar{11}}^2=-\done,\quad 2 R_2A_{\bar{01}}^1 + 2 R_2A_{\bar{11}}^1=\dtwo
\end{array}
\]
then we work out the covariant derivative using (\ref{rienabla}).
\eproof
\\

There is no regularity condition for the universal calculus on the frame group (because there is no
element different from the identity which lies outside the subset
defining the calculus).

\begin{propos}
The Riemann and Ricci tensors corresponding to the above connection are:
\[\begin{array}{l}
{\rm Riemann}(e_1)=-2\bigl(- \del^2 a +\done(\Theta_1-R_2 a)+\frac{\done \dtwo}{2}\bigr)e_1 \wedge e_2 \tens e_1\\
{\rm Riemann}(e_2)=-2\bigl(\dtwo(R_1 b-\Theta_2)+ \del^1 b -\frac{\dtwo \done}{2}\bigr) e_1 \wedge e_2 \tens e_2
\end{array}
\]
\[\begin{array}{ll}
{\rm Ricci}=&\bigl(-\del^2 a +\done(\Theta_1-R_2 a)+\frac{\done \dtwo}{2}\bigr) e_1 \otimes e_2 \\
&-\bigl(\dtwo(R_1 b-\Theta_2)+ \del^1 b -\frac{\dtwo \done}{2}\bigr)
e_2 \otimes e_1
\end{array}
\]
\end{propos}
\proof We have:
\[\begin{array}{l}
F_{\bar{10}}=\extd A_{\bar{10}}+A_{\bar{01}} \wedge A_{\bar{11}} + A_{\bar{11}} \wedge A_{\bar{01}} -2 A_{\bar{10}} \wedge A_{\bar{10}} -A_{\bar{10}} \wedge A_{\bar{01}}\\
\quad\quad  -A_{\bar{01}} \wedge A_{\bar{10}}-A_{\bar{10}} \wedge A_{\bar{11}}-A_{\bar{11}} \wedge A_{\bar{10}}\\
F_{\bar{01}}=\extd A_{\bar{01}}+A_{\bar{10}}\wedge A_{\bar{11}}+ A_{\bar{11}} \wedge A_{\bar{10}}-A_{\bar{10}}\wedge A_{\bar{01}}- A_{\bar{01}} \wedge A_{\bar{10}}\\
\quad\quad  -2 A_{\bar{01}} \wedge A_{\bar{01}} -A_{\bar{01}} \wedge A_{\bar{11}} -A_{\bar{11}} \wedge A_{\bar{01}}\\
F_{\bar{11}}=\extd A_{\bar{11}}+A_{\bar{10}} \wedge A_{\bar{01}}+A_{\bar{01}} \wedge A_{\bar{10}}-A_{\bar{11}} \wedge A_{\bar{10}} -A_{\bar{10}}\wedge A_{\bar{11}} \\
\quad\quad -A_{\bar{11}} \wedge A_{\bar{01}} -A_{\bar{01}} \wedge A_{\bar{11}}
\end{array}
\]
and
\[
{\rm
Riemann(e_1)}=-2(F_{\bar{10}}+F_{\bar{11}})\otimes e_1,\
{\rm Riemann(e_2)}=-2(F_{\bar{01}}+F_{\bar{11}})\otimes
e_2
\]
The same result is obtained by ${\rm Riemann}(e_a)=((\id \wedge \nabla)-(\extd \otimes \id))\circ \nabla (e_a)$. 
\eproof
\begin{propos}
The condition $\nabla \eta=0$ is satisfied if and only if
$\alpha=-\gamma$ and $\beta=-\delta$ and $\done=\dtwo=0$. In this case, both the Riemann
and the Ricci tensor are zero.
\end{propos}
\proof The first part of the proposition is easily proved by
computing
\[\begin{array}{ll}
\nabla (e_1 \tens e_1 +e_2 \tens e_2)=
&4a e_1 \tens e_1 \tens e_1 -2 \done e_2 \tens e_1 \tens e_1\\
&-2 \dtwo e_1 \tens e_2 \tens e_2 +4 b e_2 \tens e_2
\tens e_2=0
\end{array}
\]
which means $a=b= \done=\dtwo=0$ have
; this implies that the Riemann and the Ricci tensor are both
zero. \eproof
\\
\subsubsection{2D calculus on $\Z_2\times\Z_2$}
The calculus on the fibre will be defined now by $\{\bar{10},\bar{01}\}$
\begin{propos}
the moduli space of torsion free, cotorsion free connections is
given by:
\[
\begin{array}{l}
A_{\bar{10}}=\alpha e_1 -\frac{\partial^1 \Theta_2}{2} e_2\\
A_{\bar{01}}=-\frac{\partial ^2 \Theta_1}{2} e_1 + \beta e_2
\end{array}
\]
We set $a=\alpha,b=\beta$    (as in the case before but with
$\gamma=\delta=0$), then the  covariant derivative has the same form
as in Proposition~4.8. The regularity condition is
\[
 a \bar\del^1 b- b\bar\del^2 a=0.\]
\end{propos}
\proof We solve the torsion equations
\[
\begin{array}{l}
2 A_{\bar{10}}^2=-\done,\quad -2 A_{\bar{01}}^1=\dtwo
\end{array}
\]
and the cotorsion equations
\[
\begin{array}{l}
-2 R_1A_{\bar{10}}^2=-\done,\quad 2 R_2A_{\bar{01}}^1=\dtwo
\end{array}
\]
The regularity condition is, in this case, $A_{\bar{10}} \wedge A_{\bar{01}} + A_{\bar{01}}
\wedge A_{\bar{10}}=0$, which comes out as $a R_1 b-b R_2 a=0$, which can be written as stated.  \eproof
\\
\begin{corol}
The Riemann and Ricci tensors are (as functions of $a,b$) as in
Proposition 4.9
\end{corol}
\proof This follows from $\nabla$  but can also be computed directly as useful check; the curvature two form corresponding
to the regular connection above, is given
by:
\[\begin{array}{l}
F_{\bar{10}}=(-\Theta_2 \bar \del^2 a +\done(\Theta_1-R_2 a)+\frac{\done \dtwo}{2})e_1 \wedge e_2\\
F_{\bar{01}}=(\dtwo(R_1 b-\Theta_2)+\Theta_1 \bar \del^1 b -\frac{\dtwo \done}{2})e_1
\wedge e_2
\end{array}
\]
computed from the expression for $F$ (regularity condition applied)
\[\begin{array}{l}
F_{\bar{10}}=\extd A_{10}-2A_{10} \wedge A_{10}\\
F_{\bar{01}}=\extd A_{01}-2 A_{01} \wedge A_{01}
\end{array}
\]
the Ricci tensor is $F_{\bar{01}} e_1\tens e_2 - F_{\bar{10}} e_2 \tens
e_1$, Riemann is given by ${\rm Riemann}(e_1)=-2 F_{\bar{10}}\otimes e_1,\ {\rm Riemann}(e_2)=-2F_{\bar{01}}\otimes e_2$.
\eproof \\
Finally, the only connection fulfilling the condition
\begin{eqnarray*}
\nabla \eta&= &\nabla(e_1 \tens e_1 +e_2 \tens e_2)\\
&=&-4 a e_1\tens e_1 \tens e_1 +2 b e_2\tens e_2 \tens e_2 
+\done (e_1 \tens e_2 \tens e_1+e_1 \tens e_1 \tens
e_2)\\ & &\quad -2 \dtwo e_1 \tens e_2 \tens e_2\\ &=&0
\end{eqnarray*}
is, in this case, the null connection.

We see again the  same phenomenon as in Section~4.2; working with the 'correct' 2D calculus rather than
the universal 3D eliminates redundant fields that do not enter into the Riemannian geometry, 
trading them for an optional regularity condition. 

\subsection{Discrete sphere base with $\Z_3\subset SO(2)$ frame group}

As the main alternative to the above models, we look at the case of the universal calculus on the 4 points of our base space, which has the connectivity
of a tetrahedron or discrete model of a sphere:
\[
\includegraphics[scale=0.3]{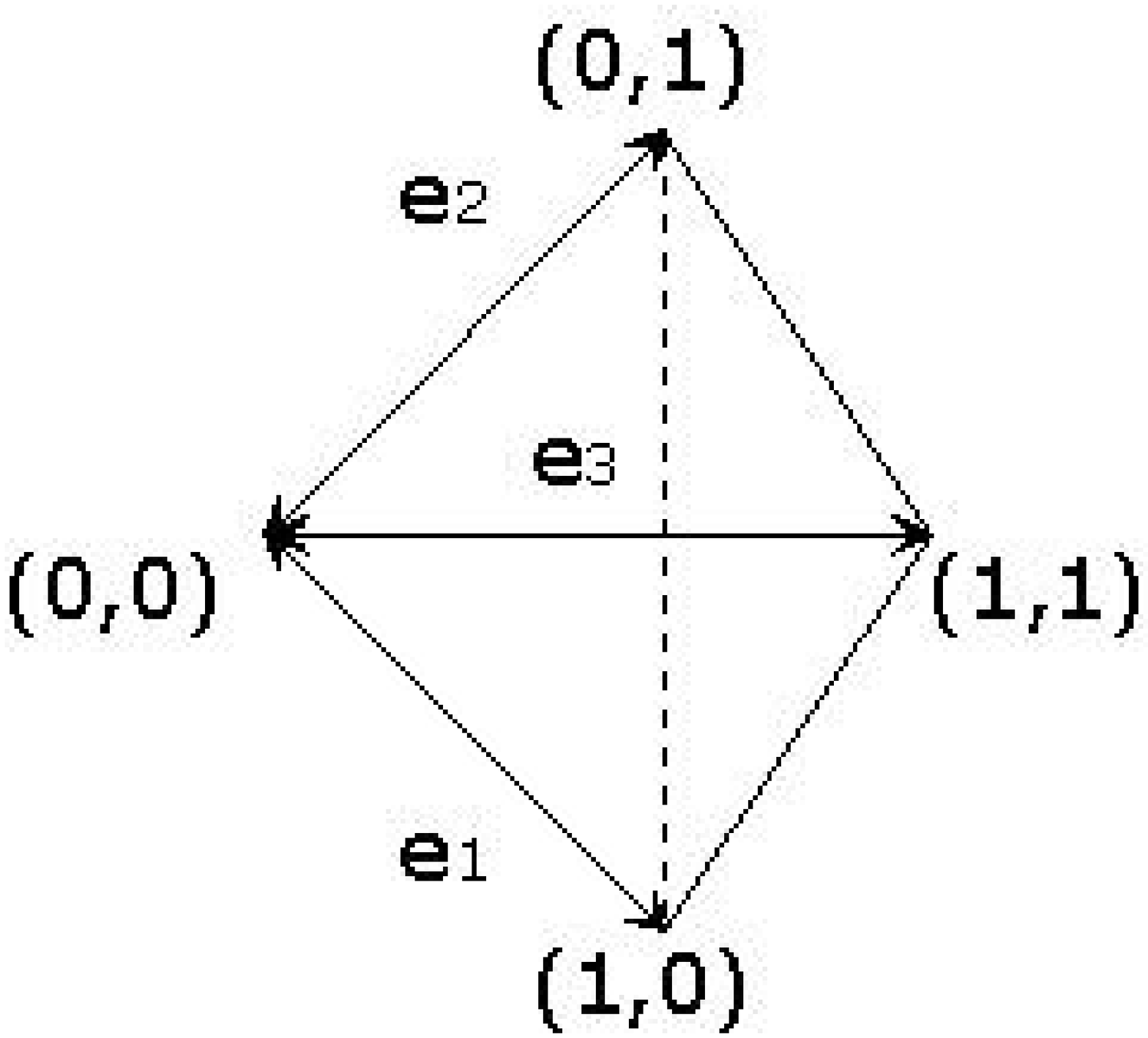}
\]
Our results are rather unusual, probably due to the small number of points in the model. As a projector we are led to $\pi$ defined by
\[
\pi(e_a \otimes e_b)=i(e_a \wedge e_b)=
\begin{cases}
0 &  a \neq b\\
\frac{1}{3}(e_1 \otimes e_1 +e_2 \otimes e_2 +e_3 \otimes e_3) & a=b
\end{cases}.
\]
This means that $\Omega^2$ has the relations
\[ e_1^2=e_2^2=e_3^2\equiv {\rm Top},\quad e_a\wedge e_b=0,\quad\forall a\ne b.\]
This projector obeys the compatibility condition (i) of Theorem 2.2 as follows.  We are required to list all the two-arcs contained in the graph.  Naming the vertices as $x,y,z,t$ (starting from $(0,0)$ and going clockwise)
the possible two arcs from $x$ are:
\[\begin{array}{c}
x\rightarrow y\rightarrow x,
x\rightarrow y\rightarrow z,
x\rightarrow y\rightarrow t\\
x\rightarrow z\rightarrow x,
x\rightarrow z\rightarrow y,
x\rightarrow z\rightarrow t\\
x\rightarrow t\rightarrow x,
x\rightarrow t \rightarrow y, 
x\rightarrow t\rightarrow z
\end{array}
\]
Now we have to make sure all the posible expression of the form
\[{\pi_x}_{yx}^{yz},
{\pi_x}_{yx}^{yt},
{\pi_x}_{yx}^{zy},
{\pi_x}_{yx}^{zt},\dots
\]
(60 of them in total) vanish, which happens to be the case. Note that we just considered the two arcs departing from $x$, since the choice of "start point" is immaterial here due to the symmetry of the graph.
The second condition of Theorem~2.2 is empty in this case, because the calculus  on $\Sigma$ is the  universal one.
$\Omega^2(\Sigma)$ 
The action of the external derivative on the vielbein elements $e_i$ is computed from~\ref{rhodel} and is
\[\extd e_1=(\Theta_1+R_1 \Theta_1){\rm Top}\equiv \tilde{\Theta_1}{\rm Top},\quad \extd e_2=(\Theta_2+R_2 \Theta_2){\rm Top}\equiv\tilde{\Theta_2}{\rm Top}\]
\[ \extd e_3=(\Theta_3+R_3 \Theta_3){\rm Top}\equiv \tilde{\Theta_3}{\rm Top}\]

Next,  we take $\Z_3=\{\bar{0},\bar{1},\bar{2}\}$ as a frame group, with calculus defined by
$\{\bar{1},\bar{2}\}$. 
$f^{\bar{1}},f^{\bar{2}}$ will acting on $e_1,e_2,e_3$ as $e_1\rightarrow e_3 \rightarrow e_2 \rightarrow e_1$ (notice that the definition of the projector is invariant under this action), or:
\[
\begin{array}{ll}
f^{\bar{1}} \triangleright e_1=e_3-e_1, & f^{\bar{2}} \triangleright e_1=e_2-e_1\\
f^{\bar{1}} \triangleright e_2= e_1 -e_2, & f^{\bar{2}} \triangleright e_2=e_3-e_2\\
f^{\bar{1}} \triangleright e_3= e_2 -e_3, & f^{\bar{2}} \triangleright e_3=e_1-e_3
\end{array}
\]
(it's an anticlockwise rotation in the picture below, which is the tetrahedron from the viewpoint of the vertex $(0,0)$)
\[
\includegraphics[scale=0.3]{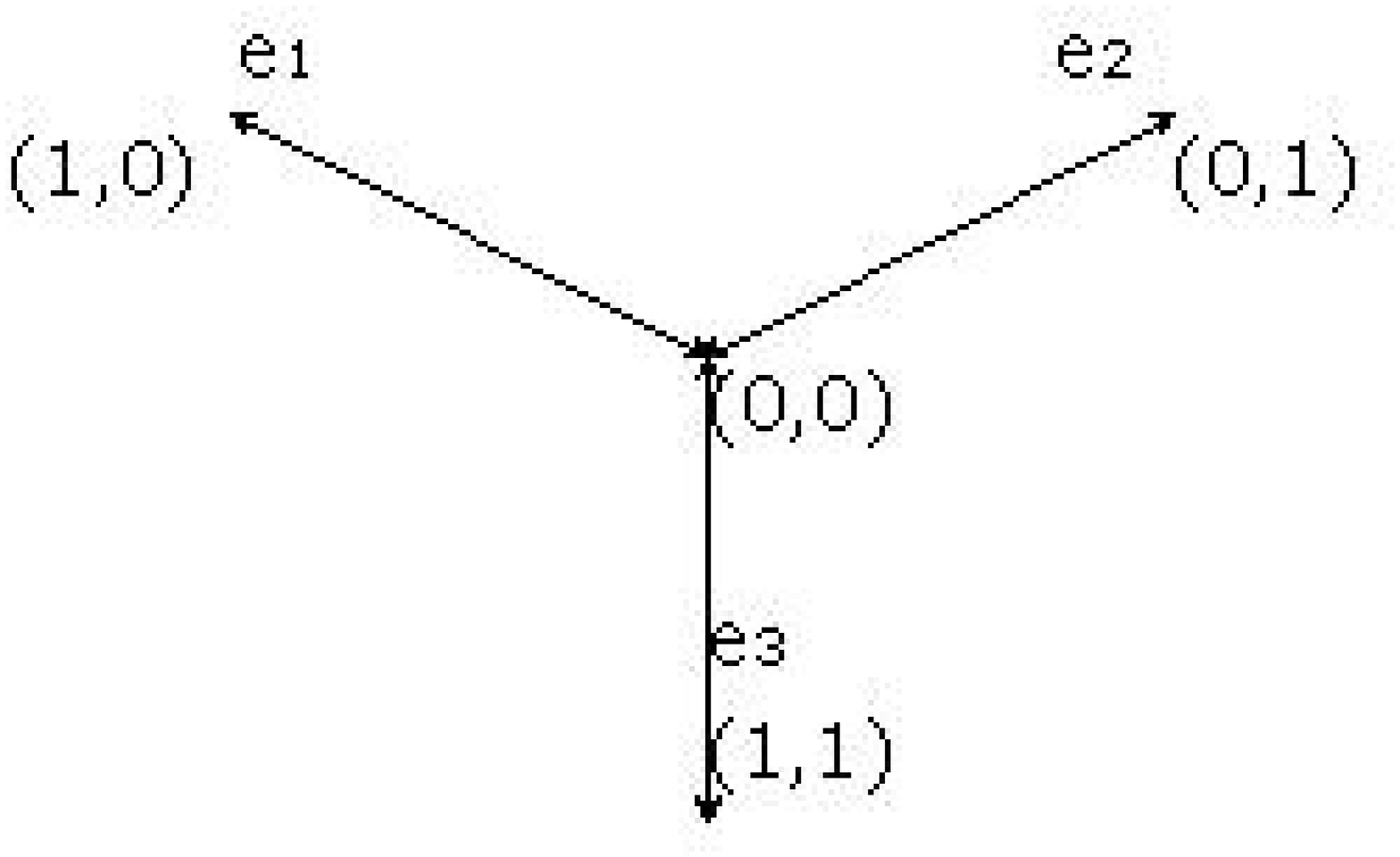}
\]
\begin{propos}
The moduli space of  torsion free connections is 4-dimensional, given by 6 parameters
$\alpha_1,\dots,\alpha_3,\beta_1,\dots,\beta_3$ with two independent equations given by
\[\tilde{\Theta_1}+\alpha_3-\alpha_1-\beta_1+\beta_2=0\]
and cyclic permutations.
Tha additional conditions for zero-cotorsion are the  two independent equations given by  
\[\bar \partial^1 (\alpha_1+\beta_1)-\bar \partial^2 \alpha_2 -\bar \partial^3 \beta_3 -\alpha_2+\alpha_3+\beta_2 -\beta_3=0\] and cyclic permutations.
\end{propos}
\proof
Firstly, we write down the zero torsion condition, but with a notation of the form
\[A_1^1=\alpha_1,\ A_1^2=\alpha_2,\ A_1^3=\alpha_3,\ A_2^1=\beta_1,\ A_2^2=\beta_2,\ A_2^3=\beta_3\] to underline a symmetry of the theory with respect to cyclical permutations in the upper indexes of the $A_i^j$ (as usual, the lower index refers to the frame group directions).
The vanishing of the cotorsion corresponds to
\[\tilde{\Theta_1}+R_2 \alpha_2-R_1 \alpha_1+R_3 \beta_3 - R_1 \beta_1=0\] and cyclic permutations.
Combining the torsion and cotorsion equations we  obtain the equations as stated. 
\eproof \\
The covariant derivative shows the same rotational symmetry.
Infact, given 
\begin{eqnarray*}
\nabla e_1&=&(\omega+\tilde\omega)\tens e_1-\omega\tens e_2-\tilde\omega\tens e_3\\
\omega&=&\sum_a\alpha_a e_a,\quad \tilde\omega\ =\ \sum_a\beta_a e_a
\end{eqnarray*}
$\nabla e_2$ and $\nabla e_3$ can be found by cyclical rotations of $\nabla e_1$.
\begin{propos}
The Riemann tensor corresponding to the connection above is
\[{\rm Riemann}(e_1)=-(\rho+\tilde{\rho}){\rm Top} \otimes e_1+ \tilde{\rho}{\rm Top} \otimes e_2 +\rho {\rm Top}\otimes e_3\]
(${\rm Riemann}(e_2), {\rm Riemann}(e_3)$ can be found by cyclical rotation)
where
\begin{eqnarray*}
\rho=\partial^1 \alpha_1+\alpha_1 \tilde{\Theta_1}+\beta_1R_1(\beta_1-\alpha_1)-\alpha_1R_1(2 \alpha_1+\beta_1)+\mbox{cycl.}\\
\tilde{\rho}=\partial^1 \beta_1+\beta_1 \tilde{\Theta_1}+\alpha_1R_1(\alpha_1-\beta_1)-\beta_1R_1(2 \beta_1+\alpha_1)+\mbox{cycl.} \
\end{eqnarray*}
and the Ricci scalar, $R=-(\rho+\tilde{\rho})$
\end{propos}
\proof
Riemann tensor is obtained in the usual way from the curvature components $F_i$; the expression for Ricci then comes out as
\[
\begin{array}{ll}
{\rm Ricci}=&\frac{1}{3}[-(F_1+F_2)e_1 \otimes e_1+F_2 e_1 \otimes e_2+F_1 e_1 \otimes e_3
+F_1 e_2 \otimes e_1
-(F_1+F_2)e_2 \otimes e_2\\
&+F_2 e_2 \otimes e_3+F_2 e_3 \otimes e_1+F_1 e_3 \otimes e_2 -(F_1+F_2)e_3 \otimes e_3]
\end{array}
\]
(from which the Ricci scalar $R=-(F_1 + F_2)$). The curvature two-form is:
\[
F_1=dA_1+A_2 \wedge A_2-2 A_1 \wedge A_1 -A_2 \wedge A_1 - A_1 \wedge A_2, F_2=dA_2+A_1 \wedge A_1 -A_1 \wedge A_2 -A_2 \wedge A_1 -2 A_2 \wedge A_2
\]
and similarly for the other components. 
\eproof

We know that the moduli space of connections is 2-parameter, which we see here is reflected in the  two physical curvature parameters $\rho, \tilde\rho$.  This model is obviously far from classical, but we see that it has several reasonable features including a cyclic symmetry and a degree 2 top form, i.e. a nonclassical 'surface'. 

\section{Remarks on the quantum theory}

So far we have solved only for the classical geometry which could form the
basis for classical equations of motion for gravity and matter in a classical background.
For quantum theory at least in a path integral approach one must integrate over
all such moduli spaces with respect to an action weighting. Here quantum gravity, in
particular, diverges badly. The advantage of working only on a finite number of points
as we have done above is that now such functional integrals become finite dimensional
integrals, which may still diverge but which are surely much more tractable.
Such integrals for gauge theory on $S_{3}$  are discussed in \cite{MaRai:ele}  and carried to
fruition for Yang-Mills on $\Z_{2}\times\Z_{2}$ in \cite{Ma:clif}, where the theory was
found to be divergent but renormalisable. Here we make some first remarks about how to extend this
in principle to the gravitational case.  The new ingredient not yet covered is the correct
'unitarity'  or reality conditions on the spin connection, which we now propose.

Thus, until now we could have worked above over a generic field, but now we must really
 we work over $\C$ and 
specify reality or `unitarity' conditions which should be expected
for a physical interpretation. This cuts down our moduli still
further  and also reduces us to integration over real variables in our finite setting.
To do this, we note that $\C(\Sigma) $  is a $*$-algebra with $*$ given by pointwise
complex conjugation. We extend this to inner calculi with the assumption $\theta^{*}=\theta $ so that $*$ anticommutes with $\extd$ (other conventions are also possible). For models based on
groups and conjugacy classes with elements of order 2 this is naturally implemented by $e_{a}^{*}=e_{a}$ (more generally, $e_{a^{-1}}$), as in \cite{MaSch,MaRai:ele,Ma:clif}.
For the models based on $\Z_2\times\Z_2$ connectivity in Section~4 we take $e_a^*=e_a$. 
We likewise, and more importantly, we take 
\[ \theta^*=\theta\]
 which ensures that $\extd=[\theta,\ \}$ 
behaves as usual for a $*$ structure in the differential graded algebra (so $\extd$ graded-commutes with $*$). In terms of field components 
this translates to  
\[ \overline{\Theta_a(x)}=R_a \Theta_a(x).\]
This is consistent with our model in Section~4, for example, where the condition (\ref{constraint})  on $\Theta$ in Section~4.1 for a 2-form projector is invariant under $*$. 

Next, we consider the spin connection components. For a unitary action for the braided-Lie algebra generators $f^i$ we would take $A_i^*=A_i$. What is a unitary action is motivated from Hopf algebra theory where the action of a Hopf $*$-algebra one would require $(f^i\la e_a)^*=S^{-1}(f^i{}^*)\la e_a^*$, where in our case $Sf^i=f^{i^{-1}}$ is inversion in the frame group algebra. The $*$-structure on the braided-Lie algebra generators which is not so clear, but if we assumed that
$f^i{}^*=f^{i^{-1}}$ as for elements in a group algebra, these two inverses cancel and we would be
led to  require $(f^i \triangleright e_a)^*=f^i \triangleright e_a$. This indeed holds for  the actions in the present paper, particularly those in Section~4, since these are obtained from permutations. Next, if the generators are unitary in this sense, we want the frame group connection to be 'antihermitian' so we propose here  
\[ A_{i}^*=A_{i^{-1}}.\]
for the component 1-forms. This has the reasonable consequence that applying $*$ to the torsion equations gives the cotorsion equations, i.e. these are related by complex conjugation in the unitary version of the theory. This is desirable as it suggests that imposing the unitarity condition on the moduli space of torison and cotorsion free connections is not so likely to give no solutions.  This too is bourne out when we look closely at the moduli of connections on our $\Z_2\times\Z_2$ in Proposition~4.2 or~4.5. We concentrate on the second of these as the more physical model with modes $a,b$.

\begin{propos} The reality condition in the moduli of torsion free and cotorsion free connections on the discrete torus in Proposition~4.5 is $\bar a=R_2 a$, $\bar b=R_1 b$. The regularity condition is invariant under conjugation and the Ricci scaler in Proposition~4.5 is real up to a `total divergence' given by $\bar\del^1,\bar\del^2$. 
\end{propos}
\proof From the above, we deduce from Proposition~4.5 and the reality condition on the $\theta$, we find $\bar\alpha=-R_1\beta$, $\bar\beta=R_2\alpha$ which translates as stated given that the functions
$a,b$ reverse sign under $R_1R_2$. The latter also means that $R_1(\bar\del^2 a)=\bar\del^2(-R_2 a)=\bar\del^2 a$, and similarly $\bar\del^1b$ is $R_2$-invariant. Since $(\bar\del^2\theta_1)^*=R_1(\bar\del^2\theta_1)$, and similarly with $R_2$ for $\bar\del^1\theta_2$, we see that the regularity condition is invariant under $*$. We then compute
\begin{eqnarray*}\bar R&=&R_1(\del^1 b+\bar\del^2\theta_1 R_1b)+R_2(\del^1a+\bar\del^1\theta_2 R_2a)-2b R_1 b- 2a R_2 a\\
&=& R+\bar\del^1(\del^1b+\bar\del^2\theta_1 R_1b)+\bar\del^2(\del^2a +\bar\del^1\theta_2 R_2a).
\end{eqnarray*}
where $(\del^1 b)^*=(R_1\theta_1)\del^1 R_1 b=R_1(\del^1 b)$, and similarly for $\del^2 a$.
\eproof

The reduced moduli space with full metric compatibility in Proposition~4.7 is also consistent with this $*$-structure, i.e. our reality condition holds for $a,b$ given by $\theta_a$ as stated there. Moreover, the stated condition on the $\theta_a$ required for this reduces simply to $a,b$ real. 
 
After that, for quantum gravity one should presumably take as action $S=\sum_{x\in\Sigma} R(x)$ using the
Ricci scalar curvature; we are not in a position to deduce field equations by a variational principle, so
this is an assumption of one way to make sense of the quantum theory. To see how this works we again
look at our discrete torus model on 4 points. 
We already know from Section 3 that for 2 or 3 points the Ricci scalar vanishes in all our models, so this model would be the first with nontrivial Ricci scalar.  From the above Propostition~5.1 we see that the action $S$ is real. Moreover, our fields $a,b$ etc are functions on the four points but so highly constrained as to be fully determined each by a single complex number, which we denote $A,B$. Here
\[ A=a(0,0)=-a(0,0),\quad \bar A=a(0,1)=-a(1,0)\]
\[  B=b(0,0)=-b(1,1),\quad \bar B=b(1,0)=-b(0,1).\]
Note that the Ricci scaler splits up into two terms
\[ R=R_B+R_A;\quad R_B=\del^1 b+\bar\del^2\theta_1 R_1b-2b R_1 b=R_2\theta_1\bar b-\theta_1 b-2b\bar b\]
and the similar expression for $R_A$ with $1,2$ interchanged. Writing 
\[ \Theta=\theta_1(0,0), \quad \bar\Theta=\theta_1(1,0),\quad \tilde\Theta=\theta_1(0,1),\quad\overline{\tilde\Theta}=\theta_1(1,1)\]
we find
\[ S=S_B+S_A;\quad S_B=-8 B\bar B+2 B(\overline{\tilde\Theta}-\Theta)+2\bar B(\tilde\Theta-\bar\Theta).\]
where we compute $R_B$ at the four points in terms of our new variables and add up. Similarly for the $A$ field and $\theta_2$. If we restrict to the full metric compatibility in Theorem~4.4 then the action is just $S_B=-4B^2$ and the dynamical variables are $\Theta,\tilde\Theta$ constrained such that  $B=\tilde\Theta-\bar\Theta$ is real. Again similarly for the $A$ system. 

Finally, we make a polar decomposition of the fields as
\[ B=\lambda e^{\imath\phi},\quad \Theta=\mu e^{\imath\psi},\quad \tilde\Theta=\tilde\mu e^{\imath\tilde\psi}\]
in terms of real positive $\lambda,\mu,\tilde\mu$ and angles $\phi,\psi,\tilde\psi$. In terms of these, we find
\[ S_B=-8\lambda^2+4\lambda\tilde\mu \cos(\phi-\tilde\psi)-4\lambda\mu\cos(\phi+\psi)\]
 with similar results for the $A$ system. Then 'quantum gravity' is reduced to integrals over these real  variables.  There remains the constraint (\ref{constraint}) as well as the optional  regularity condition to be imposed on the moduli in Proposition~4.5. These both cross-couple the $A$ and $B$ systems making even this simplest model nontrivial.  
 
 It is not our scope to consider the quantum theory in detail here,  particularly since the geometries in this paper are low dimensonal, where one does not expect very dynamical quantum gravity; for a compact surface in two dimensions the integral of the classical Ricci scalar is a constant by the Gauss-Bonnet theorem.  For a classical torus this should be zero, so we see that the discrete torus model already exhibits non-standard behaviour, the meaning of which remains to be understood. It also remains to identify physical observables to be computed by such functional integral methods. However, our low-dimensional example does indicate the possibility of reasonable unitarity constraints and illustrate how a quantum gravity theory might proceed in principle.

\bigskip

\section{Combinatorics of geometries up to nine points}

For higher numbers of points we do not attempt a detailed
classification but rather we overview the range of possibilities
with a view to picking out the most interesting ones.

In the first place, we now limit ourselves to the  more
interesting case of symmetric (`bidirectional') differential
calculi. These are just graphs with no self-edges and no more than
one edge between vertices. For a fibration with fiber size $n$,
these are the so-called $n$-regular graphs. There is no
classification theory for $n$-regular graphs (eg any $n$-regular
simplicial approximation of a manifold gives one) but small ones
are listed in \cite{ReaWil}. From there we see that there is a
reasonable number for $m\le 8$ after which the number grows
rapidly. We deal only with connected graphs.

Note also for any $m$ that here are none for $n=1$ (except $m=2$).
For $n=2$ there is just the $m$-gon for all $m$. This is the
differential calculus on $\Z_m$ with $\CC=\{-1,1\}$. For $n=m-1$
there is exactly the universal calculus or totally connected
graph. We observe that the $m$-gon and universal calculi are
members of a `circulant graph' family $\Z_m{}^{(1,p,q,\cdots)}$
where $p,q,\cdots$ are distinct integers modulo $m$. They
correspond to the calculus on $\Z_m$ with $\CC=\{\pm1,\pm p, \pm
q,\cdots \}$ where we only have $p$ if $2p=0$ mod $m$, etc. The
direct product of circulants with $\CC_1,\CC_2$ means with
$\CC=(\CC_1,0)\cup(0,\CC_2)$ (as for the product of any groups
equipped with differential structures, see \cite{Ma:rief}). An
example of a circulant is in Figure~1. Note also the `handshaking
lemma' in graph theory that $nm$ has to be even. Then we have the following list of
connected graphs which is complete up to $m=8$:

{\bf For $m=2$} we have only the universal calculus at $n=1$.

{\bf For $m=3$} we have only the universal calculus which equals
the 3-gon calculus at $n=2$.

{\bf For $m=4$} we have only the 4-gon at $n=2$, which can also be
viewed as $\Z_2^{(1)}\times \Z_2^{(1)}$ (i.e. with the direct
product calculus where $\CC=\{(0,1),(1,0)\}$), and the universal
calculus at $n=3$.

{\bf For $m=5$} we have only the 5-gon at $n=2$ and the universal
at $n=4$.

{\bf For $m=6$} we have only the 6-gon at $n=2$ and two choices at
$n=3$. These are the circulant $\Z_6^{(1,3)}$, which is also the
graph for the $S_3$ calculus with its 2-cycles conjugacy class,
and the circulant $\Z_2^{(1)}\times\Z_3^{(1)}$. At $n=4$ we have
only the circulant $\Z_6^{(1,2)}$, which is a triangulation of the
sphere and is also the graph for $S_3$ with a left-covariant
calculus. See Fig 1 (a),(b). At $n=5$ we just have the universal
one. Note that the 3-cycles calculus on $S_3$ is not connected so
does not appear in this list.

{\bf For $m=7$} we have only the 7-gon at $n=2$, none at $n=3,5$
and two choices at $n=4$. One is the circulant $\Z_7^{(1,2)}$ and
the other is shown in Fig 1 (c). At $n=6$ we just have the
universal one.

{\bf For $m=8$} we have the 8-gon at $n=2$ and five at $n=3$. One
of these is the cube, which is
$\Z_2^{(1)}\times\Z_2^{(1)}\times\Z_2^{(1)}$. It can also be
viewed as $\Z_2^{(1)}\times\Z_4^{(1)}$. Another is the circulant
$\Z_8^{(1,4)}$. See Fig. 1(b). The remaining three are as in Fig.
1(c). At $n=4$ there are six, namely the circulants $\Z_8^{(1,2)}$
and $\Z_8^{(1,3)}$, and $\Z_2^{(1)}\times\Z_4^{(1,2)}$ and the
remaining three in Fig. 1(c). At $n=5$ there are three, namely the
circulants $\Z_8^{(1,2,4)}$ and
 $\Z_8^{(1,3,4)}$ and the remaining one in Fig~1(c).
At $n=6$ we have only the circulant $\Z_8^{(1,2,3)}$. At $n=7$ we
just have the universal calculus.

{\bf For $m=9$} there is the $9$-gon at $n=2$, none at $n=3$ and
already sixteen at $n=4$, of which three are groups, namely the
circulants $\Z_9^{(1,3)}$, $\Z_9^{(1,4)}$ and a simplicial torus
torus (see Fig. 1(b)), which is $\Z_3^{(1)}\times\Z_3^{(1)}$.
There are none at $n=5$ and three at $n=6$ of which two are
circulants on $\Z_9$, and so forth. Fig. 1(d) also shows an
important $m=10$ graph with $n=3$ which is a $\Z_2$ quotient of
the dodecahedron and can be thought of as a discrete $\Bbb R \Bbb
P^2$.

\begin{figure}
\[ \epsfbox{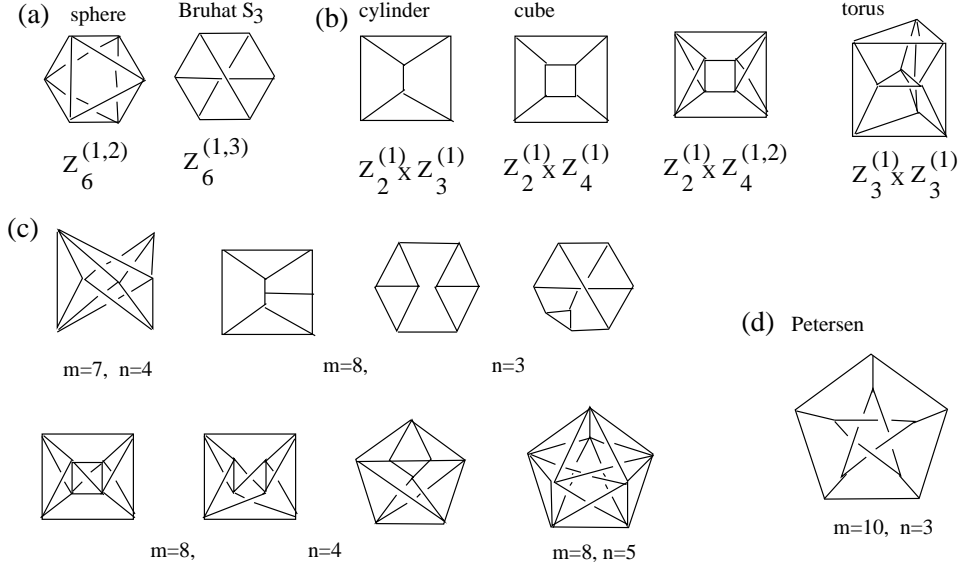}\]
\caption{(a) examples of circulant graphs (b) all products of
circulants up to $m=9$ (c) graphs up to $m=8$ not circulants or
products of them. (d) Petersen graph at $m=10$}
\end{figure}

At the qualitative or `topological' level of this section, we can
immediately present one discrete moduli space of combinatorial
solutions for vielbeins. Namely, for any $E$ that fibers over
$\Sigma$ with $|F_x|=n$, an $n$-bein is provided by any choice of
bijections $s_x:\{1,\cdots,n\}\to F_x$ by \eqn{efib}{
e_{axy}=\delta_{s_x(a),y}=e_a^{-1xy},\quad e_a=\sum_x
\delta_x\extd\delta_{s_x(a)}} giving \eqn{rhodelfib}{ e_a
f=f(s_\cdot(b))e_a,\quad (\del^a f)(x)=f(s_x(a))-f(x).} Here
$s_\cdot(a)$ is a function on $\Sigma$ (with $\cdot$ denoting the
functional dependence). Pictorially, we label all 1-arcs
arbitrarily by $\{1,\cdots,n\}$ and $s_x(a)$ is the endpoint of
the arc labelled $a$ from $x$. The element $\theta$ and the
relations of the maximal prolongation are \eqn{thetafib}{
\theta=\sum_a e_a,\quad \sum_{x{\buildrel a\over \to}{\buildrel
b\over \to}z} e_a\wedge e_b=0, \quad \forall x\ne z,\ x\to\kern
-11pt /\ \ z.} The corresponding projectors are
\eqn{pimax}{\pi(e_a\tens e_b)=e_a\tens e_b - \sum_x{\delta_x \over
|F_{x,z}|}\sum_{x{\buildrel c\over \to}{\buildrel d\over \to}z}
e_c\otimes e_d; \quad {\rm where}\ x{\buildrel a\over
\to}{\buildrel b\over\to z}} and have a functional dependence.
Since the wedge product is given by setting to zero the elements
of the tensor product which are in the kernel of this projector,
we have the lift $i:\Omega^2\to \Omega^1\tens\Omega^1$ given by
the same formula. These formulae are for general
left-parallelizable calculi. In our bidirectional case each arc
really means two arrows since we can move along it in either
direction. In this case the combinatorial data $\{s_x(a)\}$ for
this class of vielbeins is a bicolouring of the graph, with two
colours $a\in\{1,\cdots,n\}$ for each arc, namely one for each
arrow. Moveover, we can follow the coloured arrows from vertex to
vertex and in this way the doubled-up graph (in which each arc is
a pair of arrows going in opposite directions) is decomposed into
coloured loops. The loops of each colour need not be connected.

For the framed geometry one must also choose a frame group $G$
acting on the vector space $V$ spanned by the vielbeins, a
calculus on the group given by an Ad-stable subset, and projectors
$\pi$. For the combinatorial solutions above it is natural to take
$G=S_n$ acting by permuting the colours, i.e. $g\la e_a=e_{g(a)}$
for a permutation $g$. We can then take (for example) the
universal differential calculus on $S_n$ where $i\in S_n-\{e\}$ so
that there is no regularity condition to solve when we use the
braided-Lie algebra with basis $\{f^i\}$. Then the torsion and
cotorsion equations for $A_i$ are linear and hence determined by
linear algebra. More generally, our choice of frame group and
associated structures have to be chosen according to what geometry
we want to model. I.e. for each choice of regular graph for the
`topology' of the finite set, we have further choices for the
actual geometry we want to model. We have already seen how this
goes for a small number of points; there are progressively more
choices as the number of points increases.

Also, for the quantum theory on should sum over all topological
configurations, i.e. graphs and colourings, and then integrate
over all moduli spaces for each colorung (eg of the restricted
variety as we have done in Section~5), weighted with some action
such as the Einstein-Hilbert one. In this way one arrives in
principle at a quantum gravity theory in which differential
structures (which goes into the graph) are summed over as well as
an additional variable. Let us note here a remarkable duality: the
sum over all coloured graphs, which is the combinatorial part of
our theory, is in the spirit of a Feynman diagram, i.e. in some
sense the discrete quantum gravity theory is somewhat like a
scalar theory in usual flat space (with $\phi^n$ interaction if we
look at $n$-regular graphs). If one wanted to take this further,
one should sum over the number of points $m$, i.e. take all finite
sets with $n$-regular graphs or the $n$-dimensionality of the
non-commutative manifold structures fixed.

\baselineskip 14pt

\end{document}